\documentclass[aps,preprint,supberscriptaddress,showpacs,preprintnumbers,amsmath,amssymb,nofootinbib,longbibliography]{revtex4-1}
\usepackage[colorlinks=true, pdfstartview=FitV, linkcolor=red, citecolor=blue, urlcolor=black, pdftitle={},pdfauthor={Tomoya Hayata, Yoshimasa Hidaka, Masaru Hongo, Toshifumi Noumi},pdfsubject={}, pdfkeywords={relativistic hydrodynamics, quantum field theory}]{hyperref}

\usepackage{amsmath,amssymb,bm}
\usepackage{graphicx}

\usepackage[normalem]{ulem}  

\newcommand{\tr}{\mathop{\mathrm{tr}}}
\newcommand{\hH}{\hat{H}}
\newcommand{\hK}{\hat{K}}
\newcommand{\hT}{\hat{T}}
\newcommand{\hU}{\hat{U}}
\newcommand{\hrho}{\hat{\rho}}
\newcommand{\ptc}[1]{{\bar{#1}}}
\newcommand{\hSigma}{\hat{\Sigma}}
\newcommand{\hS}{\hat{S}}
\newcommand{\hJ}{\hat{J}}

\newcommand{\hc}{\hat{c}}
\newcommand{\hcurrent}{\hat{\mathcal{J}}}

\newcommand{\hO}{\hat{\mathcal{O}}}
\newcommand{\hs}{\hat{s}}
\newcommand{\he}{\hat{e}}
\newcommand{\hp}{\hat{p}}

\newcommand{\hProj}{\hat{\mathcal{P}}}

\newcommand{\hpi}{\hat{\pi}}

\newcommand{\hn}{\hat{n}}
\newcommand{\px}{\ptc{x}}
\newcommand{\pt}{\ptc{t}}
\newcommand{\timefunc}{\pt}

\newcommand{\dilatation}{\sigma}

\newcommand{\delt}{\covDer_\pt}
\newcommand{\delQ}{\tilde{\delta}}
\newcommand{\covDer}{\nabla}
\newcommand{\hlambda}{\hat{\lambda}}

\newcommand{\average}[1]{\langle#1\rangle}
\newcommand{\averageLG}[1]{\langle#1\rangle^\text{LG}}

\begin{document}

\preprint{RIKEN-QHP-180, RIKEN-MP-111}
\author{Tomoya Hayata}
\affiliation{Department of Physics, The University of Tokyo, Tokyo 113-0031, Japan}
\affiliation{Theoretical Research Division, Nishina Center, RIKEN, %
             Wako 351-0198, Japan}
\author{Yoshimasa Hidaka}
\affiliation{Theoretical Research Division, Nishina Center, RIKEN, %
             Wako 351-0198, Japan}
\author{Masaru Hongo}
\affiliation{Department of Physics, The University of Tokyo, Tokyo 113-0031, Japan}
\affiliation{Theoretical Research Division, Nishina Center, RIKEN, %
             Wako 351-0198, Japan}
\affiliation{Department of Physics, Sophia University, Tokyo 102-8554, Japan}
\author{Toshifumi Noumi}
\affiliation{Theoretical Research Division, Nishina Center, RIKEN, %
             Wako 351-0198, Japan}
             
\date{March, 16, 2015}
\title{
Relativistic hydrodynamics from quantum field theory \\ on the basis of the generalized Gibbs ensemble method
}
\begin{abstract}
We derive relativistic hydrodynamics from quantum field theories by assuming that the density operator is given by 
a local Gibbs distribution at initial time. 
We decompose the energy-momentum tensor and particle current into nondissipative and dissipative parts, 
and analyze their time evolution in detail. 
Performing the path-integral formulation of the local Gibbs distribution, we microscopically derive 
the generating functional for the nondissipative hydrodynamics.
We also construct a basis to study dissipative corrections.
In particular, we derive the first-order dissipative hydrodynamic equations 
without a choice of frame such as the Landau-Lifshitz or Eckart frame. 

\end{abstract}
\maketitle

\section{Introduction and summary}
Hydrodynamics universally describes the spacetime evolution of charge densities of systems such as energy, momentum, and particle number \cite{Landau:Fluid}. 
It does not depend on microscopic details of systems, whose application covers branches of physics from condensed matter to high-energy physics. 
Among them is illuminating the recent success of relativistic hydrodynamics in describing the evolution of the quark-gluon plasma (QGP) created in heavy-ion collision experiments \cite{Back:2004je,Arsene:2004fa,Adcox:2004mh,Adams:2005dq,Romatschke:2007mq,Song:2010mg,Aamodt:2010pa}. 

The first-order relativistic hydrodynamic equations, that is, the relativistic version of the 
Navier-Stokes equations, which suffer from the violation of causality, have been derived 
by Eckart~\cite{Eckart:1940te} and by Landau and Lifshitz~\cite{Landau:Fluid}. 
The second-order equations, which resolve the causality problem 
by introducing a finite relaxation time, were derived first by Muller~\cite{Muller:1967zza} 
and also by Israel and Stewart~\cite{Israel:1979wp}. 
After the aforementioned success of relativistic hydrodynamics in describing  the QGP, 
a lot of work concerning the derivation of hydrodynamic equations has been progressively 
carried out, in which  the hydrodynamic equations are formulated 
based on the kinetic theory~\cite{Muronga:2006zx,Tsumura:2007ji,*Tsumura:2011cj,York:2008rr,Betz:2008me,Monnai:2009ad,*Monnai:2010qp,Van:2011yn,Denicol:2012cn,Jaiswal:2013npa}, 
the fluid/gravity correspondence~\cite{Baier:2007ix,Natsuume:2007ty,Bhattacharyya:2008jc,Hubeny:2011hd}, 
the phenomenological extension of the nonequilibrium thermodynamics~\cite{Koide:2006ef,Fukuma:2011pr}, 
and the projection operator method~\cite{Koide:2008nw,Minami:2012hs}.
Also, a significant method has recently been developed in which
the equilibrium-generating functional for the nondissipative hydrodynamics is constructed only by respecting symmetries of systems~\cite{Banerjee:2012iz,Jensen:2012jh}.

The aim of this work is to derive the dissipative relativistic hydrodynamic equations from quantum field theories.
Our approach is based on the recent development of the nonequilibrium statistical mechanics~\cite{sasa:2013}, 
which is essentially equivalent to the nonequilibrium statistical operator method~\cite{Zubarev:1979,Becattini:2014yxa}.
By performing the path-integral formulation of the Massieu-Planck functional, 
we present the first microscopic justification of the generating functional method~\cite{Banerjee:2012iz,Jensen:2012jh} for nondissipative parts. 
This enables us to justify a generalized argument by Luttinger~\cite{Luttinger:1964zz}, 
in which the spatial distribution of the temperature is interpreted as an auxiliary external gravitational potential. 
We also study the dissipative corrections to relativistic hydrodynamic equations by using our method.
Although we restrict ourselves to first-order equations in this paper,
our formulation provides a solid basis to proceed to the higher orders in the derivative expansion.

In the rest of this section, we briefly summarize our result. 
The relativistic hydrodynamic equations are based on the continuity equations:
\begin{align}
\covDer_\mu T^{\mu \nu} &=0, \label{eq:conservation2}\\
\covDer_\mu J^{\mu} &=0.\label{eq:conservation1}
\end{align}
Here $\covDer_\mu$ is the covariant derivative. 
$T^{\mu \nu}$ and $J^{\mu}$ are the energy-momentum tensor and particle current, respectively. 
They are decomposed into nondissipative and dissipative parts
\begin{align}
T^{\mu \nu} &= T_0^{\mu\nu} +\delta T^{\mu\nu},\\
J^{\mu} &= J^\mu_0 +\delta J^\mu .
\end{align}
In the leading order of the derivative expansion, 
the nondissipative terms have the form of a perfect fluid:
$T_0^{\mu\nu}= (e+p)u^\mu u^\nu +pg^{\mu\nu}$ and $J^\mu_0 =n u^\mu$.
Here $e$ denotes the energy density, $p$ the pressure,  $n$ the particle density, and $u^\mu$ the fluid four-velocity.
$\delta T^{\mu\nu}$ and $\delta J^\mu$ represent the dissipative parts.
In our formalism, the dissipative terms are given as 
\begin{align}
\delta T^{\mu\nu} &= -\frac{\zeta}{\beta} h^{\mu\nu}{h^{\rho\sigma}\covDer_\rho\beta_\sigma}- 2\frac{\eta}{\beta}h^{\mu\rho}h^{\nu\sigma}\covDer_{\langle\rho}\beta_{\sigma\rangle} , \label{eq:delTmunu}\\
\delta J^{\mu} &=- \frac{ \kappa}{\beta}h^{\mu\rho}\covDer_\rho\nu, \label{eq:delJmu}
\end{align}
in the leading order of the derivative expansion
, where $\beta^\mu = \beta u^\mu $ with the inverse temperature $\beta$,
and $\nu = \beta \mu$ with the chemical potential $\mu$.
Here, $\zeta,~\eta$, and $\kappa$ denote the bulk viscosity, the shear viscosity, and the diffusion constant, respectively, 
whose microscopic expressions are given by the Kubo formulas, Eqs.~\eqref{eq:bulk}$-$\eqref{eq:diffusion}. 
We introduced the spatial projection operator $P^\mu_\nu \equiv \delta^\mu_\nu+ v^\mu n_\nu$, and $h^{\mu\nu}= P^\mu_\rho P^\nu_\sigma g^{\rho\sigma}$, 
where $n_\mu$ denotes the normal vector for an isochronous hypersurface and $v^\mu$ the time vector with $v^\mu n_\mu = -1$. 
These spatial projection operators satisfy $P^\mu_\nu v^\nu=P^\mu_\nu n_\mu=0$ and $P^\mu_\rho P^\rho_\nu=P^\mu_\nu$.
We also defined tensors with angle brackets as the traceless symmetric projected parts, which are given explicitly as 
\begin{equation}
\begin{split}
A_{\langle\mu\nu\rangle}\equiv \frac{1}{2}P_{\mu}^{\alpha}P_\nu^{\beta}(A_{\alpha\beta}+A_{\beta\alpha})-\frac{1}{d-1}h_{\mu\nu}h^{\alpha\beta}A_{\alpha\beta},
\end{split}
\end{equation}
where $d$ is the spacetime dimension. $h_{\mu\nu}$ is a symmetric tensor and satisfies $h^{\mu\rho}h_{\rho\nu}=P^\mu_\nu$.
We emphasize here that the above constitutive relations, Eqs.~\eqref{eq:delTmunu} and \eqref{eq:delJmu}, with the Kubo formulas, Eqs.~\eqref{eq:bulk}$-$\eqref{eq:diffusion}, 
are obtained without choosing any frame; this is an advantage of our new formulation.
The particular choice of $v^\mu$ and $n_\mu$ reproduces the dissipative hydrodynamic equations in the known frame. 
For example, we reproduce the Landau-Lifshitz frame if we choose $v^\mu=n^\mu=u^\mu$.

This paper is organized as follows:
In Sec.~\ref{sec:LocalThermodynamics}, we review the local thermodynamics. 
In Sec.~\ref{sec:pathIntegral}, we derive the path-integral formulation of the Massieu-Planck functional on a hypersurface.
In Sec.~\ref{sec:TimeEvolution}, we discuss the time evolution of hydrodynamic variables, and derive self-consistent equations giving constitutive relations.
In Sec.~\ref{sec:DerivativeExpansion}, we discuss the derivative expansion of the hydrodynamic equations in a frame-independent way. 
Section.~\ref{sec:Discussion} is devoted to a discussion.

\section{Local thermodynamics on a hypersurface} \label{sec:LocalThermodynamics}
In this section we discuss the local thermodynamics on a spacelike hypersurface in order to construct relativistic hydrodynamic equations in a covariant way.
In Sec.~\ref{sec:Geometric}, we first summarize geometric aspects of the spatial hypersurface used in this paper.
In Sec.~\ref{sec:LocalGibbs},
we introduce several concepts such as the local Gibbs distribution and the entropy current operator based on Refs.~\cite{Zubarev:1979,vanWeert1982133,Weldon:1982aq,Becattini:2014yxa}.
In Sec.~\ref{sec:pathIntegral}, we derive the path-integral formulation of the Massieu-Planck functional on the hypersurface.
The Lagrangian is written as that in the curved spacetime background fields,
whose metric consists of the local temperature and the fluid four-velocity.
We show that the metric has Kaluza-Klein gauge symmetry in addition to ($d-1$)-dimensional diffeomorphism invariance~\cite{Banerjee:2012iz}.

\begin{figure}[h]
\includegraphics[width=0.5\linewidth]{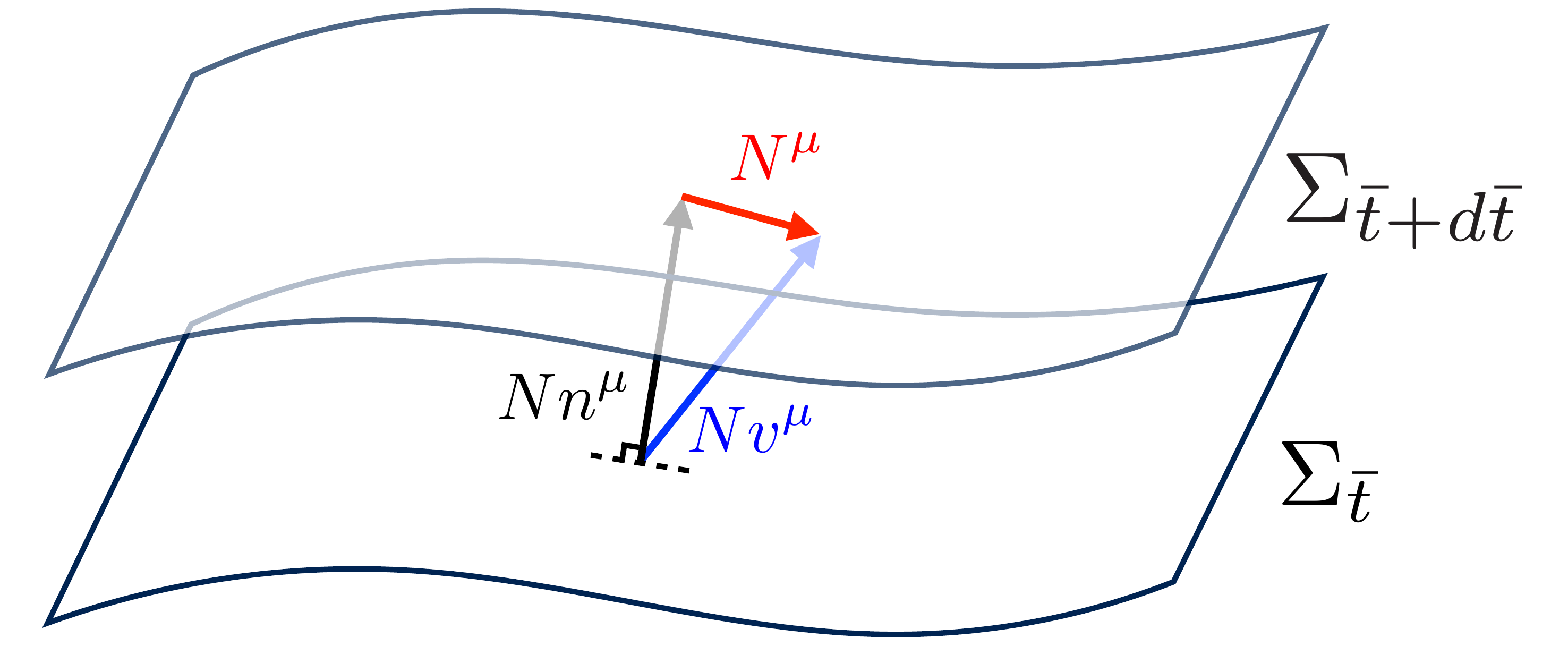}
\caption{Illustration of the Arnowitt-Deser-Misner (ADM) decomposition of the  spacetime. $\Sigma_{\pt}$ denotes a spacelike hypersurface 
parametrized by $\pt (x)=\ $const. $n^{\mu}$ is a vector normal to the hypersurface. 
Introducing the lapse function $N(x)$ and the shift vector $N^{\mu}(x)$, 
we decompose the time vector as $\partial_\pt x^\mu=Nv^{\mu} = Nn^\mu +N^\mu $.} 
\label{Fig:hypersurface}
\end{figure}

\subsection{Geometric preliminaries}
\label{sec:Geometric}
As a technical preparation,
we first summarize the geometric aspects of spacelike hypersurface in this subsection.
Let us consider spatial slicings on a general curved spacetime
with a metric $g_{\mu\nu}$
and parametrize the spacelike hypersurface by $\timefunc$.
We also introduce the spatial coordinates $\bm{\px}$
on the hypersurface.
In other words,
we define a spacelike hypersurface $\Sigma_\pt$
by the $\timefunc(x)=$ const. surface,
and introduce spatial coordinates $\bm{\px}=\bm{\px}(x)$,
where $x$ is a general coordinate (see Fig.~\ref{Fig:hypersurface}).
To discuss dynamics on such a spacelike hypersurface,
it is convenient to introduce a timelike unit vector $n_\mu$ as
\begin{align}
n_\mu(x) =-N(x)\partial_\mu \timefunc(x)
\quad
{\rm with}
\quad
N(x)\equiv (-\partial^\mu \timefunc(x) \partial_\mu \timefunc(x))^{-1/2}\,.
\end{align}
Here we normalize $n_\mu$ as $n_\mu n^\mu=-1$
and $n^\mu$ is future oriented.
$N>0$ is the lapse function.
We use the mostly plus convention of the metric,
e.g.,
the Minkowski metric is $\eta_{\mu\nu}\equiv\mathop{\mathrm{diag}} (-1,1,1,\cdots,1)$.
The induced metric $\gamma_{\mu\nu}$ on the spacelike hypersurface 
is then
\begin{align}
\gamma_{\mu\nu}=g_{\mu\nu}+n_\mu n_\nu\,.
\end{align}
We also introduce
the shift vector $N^\mu$ by
the decomposition
\begin{equation}
\label{pre_v}
\partial_\pt x^\mu(\pt,\bm{\px}) = N n^\mu+N^\mu
\quad
{\rm with}
\quad
n_\mu N^\mu=0\,.
\end{equation}
In the coordinate system $(\pt,\bm{\px} )$,
$n_\mu$, $\gamma_{\mu\nu}$,
and $N^\mu$ are given explicitly by
\begin{align}
n_{\ptc{\mu}}=(-N,\bm{0})\,,
\quad
\gamma_{\ptc{0}\ptc{i}}=\gamma_{\ptc{i}\ptc{0}}=g_{\ptc{0}\ptc{i}}=g_{\ptc{i}\ptc{0}}\,,
\quad
\gamma_{\ptc{i}\ptc{j}}=g_{\ptc{i}\ptc{j}}\,,
\quad
N^{\ptc{\mu}}= 
\begin{pmatrix}
0 \\
N^2g^{\ptc{0}\ptc{i}}
\end{pmatrix}\,.
\end{align}
The metric $g_{\ptc{\mu}\ptc{\nu}}$ takes the form of the Arnowitt-Deser-Misner (ADM) metric,
\begin{equation}
\label{ADM}
\begin{split}
g_{\ptc{\mu}\ptc{\nu}} = g_{\mu\nu} \frac{\partial x^\mu}{\partial \px^{\ptc{\mu}}}\frac{\partial x^\nu}{\partial \px^{\ptc{\nu}}}
=
\begin{pmatrix}
-N^2 +N_\ptc{i}N^\ptc{i} & N_\ptc{j}\\
N_{\ptc{i}}&  \gamma_{\ptc{i}\ptc{j}}
\end{pmatrix},\qquad
g^{\ptc{\mu}\ptc{\nu}}=
\begin{pmatrix}
-N^{-2}& N^{-2}N^{\ptc{j}}\\
N^{-2} N^{\ptc{i}}&  \gamma^{\ptc{i}\ptc{j}}-N^{-2}N^{\ptc{i}}N^{\ptc{j}}
\end{pmatrix}.
\end{split}
\end{equation}
Here $N_\ptc{i} =\gamma_{\ptc{i}\ptc{j}}N^\ptc{j}$. $\gamma^{\ptc{i}\ptc{j}}$ is the inverse of $\gamma_{\ptc{i}\ptc{j}}$ and satisfies $\gamma_{\ptc{i}\ptc{j}}\gamma^{\ptc{j}\ptc{k}}=\delta_\ptc{i}^\ptc{k}$.
The $d$-dimensional volume element  is given by
\begin{align}
\int d^dx\sqrt{-g}=\int d^dxN\sqrt{\gamma}
\quad
{\rm with}
\quad
\gamma=\text{det}\gamma_{\ptc{i}\ptc{j}}\,,
\end{align}
whereas the volume element on the spacelike hypersurface $\Sigma_\pt$ is
\begin{align}
\int d\Sigma_\pt=\int d^{d}x\sqrt{-g}\delta\big(\pt-\timefunc(x)\big)N^{-1}(x)
=\int d^{d-1}\bm{\px}\sqrt{\gamma}\,.
\end{align}
It is also convenient to introduce a vector $v^\mu$
proportional to Eq.~\eqref{pre_v} as
\begin{equation}
v^\mu=N^{-1}\partial_\pt x^\mu(\pt,\bm{\px})
\quad
{\rm with}
\quad
v^\mu n_\mu=-1
\,.
\end{equation}
Using $n_\mu$ and $v^\mu$,
we define a spatial projection operator $P^\mu_\nu$ as
\begin{equation}
\begin{split}
P^\mu_\nu \equiv \delta^\mu_\nu+ v^\mu n_\nu
\quad
{\rm with}
\quad
P^\mu_\nu v^\nu=0\,,
\quad
P^\mu_\nu n_\mu=0\,,
\quad
P^\mu_\rho P^\rho_\nu =P^\mu_\nu \,.
\end{split}
\end{equation}
Its concrete form
in the coordinate system $(\pt,\bm{\px} )$
is given by $P^{\ptc{\mu}}_{\ptc{\nu}}=\text{diag}(0,1,1,\ldots,1)$.
We will use this projection operator in Sec.~\ref{sec:TimeEvolution}.
We note that such an operator
often appears in the context of Newton-Cartan geometry
(see, e.g., Refs.~\cite{Son:2013rqa,Jensen:2014aia}~\footnote{Our normalization $n_\mu v^\mu=-1$
has the opposite sign compared to that in Refs.~\cite{Son:2013rqa,Jensen:2014aia}}).

\subsection{Local Gibbs distribution} \label{sec:LocalGibbs}
We next introduce a density operator 
representing a local thermal equilibrium state, and review the thermodynamics on the hypersurface~\cite{Zubarev:1979,vanWeert1982133,Weldon:1982aq,Becattini:2014yxa}.
We start with global thermal equilibrium on the Minkowski space, in which
the density operator  for an arbitrary inertial frame of reference
is given as the Gibbs distribution,
\begin{equation}
\begin{split}
\hrho_\text{eq}(\beta^\mu, \nu) = e^{\beta^{\mu}\hat{P}_\mu  +\nu \hat{N}-\Psi(\beta^\mu,\, \nu)},
\end{split}
\label{eq:GibbsDistribution}
\end{equation}
where parameters are $\beta^\mu=\beta u^\mu$ with the inverse temperature $\beta$, 
the fluid four-velocity of the system $u^\mu$ normalized by $u^\mu u_\mu=-1$, and
$\nu = \beta \mu$ with the chemical potential $\mu$.
$\hat{P}_\mu$ and $\hat{N}$ denote energy-momentum and number operators, respectively.
The Massieu-Planck function $\Psi(\beta^\mu,\nu)\equiv \ln \tr\exp[{\beta^{\mu}\hat{P}_\mu  +\nu \hat{N}}]$ determines the normalization of the density operator $\hrho_\text{eq}$.
At the rest frame of medium, $u^{\mu}=(1,\bm{0})$, and thus $\hrho_\text{eq}(\beta^\mu, \nu) = \exp\bigl[{-\beta(\hat{H}  -\mu \hat{N})-\Psi(\beta,\nu)}\bigr]$ are satisfied.

We then generalize the global Gibbs distribution~\eqref{eq:GibbsDistribution} to a local form in a coordinate-invariant way.
For this purpose, let us consider thermodynamics on the spacelike hypersurface, $\Sigma_\pt$, introduced in the previous subsection.
For generality, we leave the metric $g_{\mu\nu}$ of the spacetime as a general curved one.
On the hypersurface, we introduce a local Gibbs distribution $\hrho_{\text{LG}}[\pt; \lambda]$ as
\begin{equation}
\begin{split}
\hrho_{\text{LG}}[\pt; \lambda]\equiv\exp\bigl(-\hS[\pt;\lambda]\bigr)
\quad
{\rm with}
\quad
\hS[\pt;\lambda]\equiv\hK[\pt;\lambda]+\Psi[\pt;\lambda]\,,
\end{split}
\end{equation}
where $\hK[\pt;\lambda]$ is defined by
\begin{equation}
\begin{split}
\hK[\pt;\lambda] \equiv 
-\int d\Sigma_{\pt \mu}\,\lambda^a(x)\hcurrent_a^\mu(x)
=-\int d\Sigma_{\pt\nu}\, \Bigl( \beta^\mu(x){\hT^{\nu}}_{~{\mu}}(x)  +\nu(x)\hat{J}^{\nu}(x)\Bigr).
\end{split}
\end{equation}
Here we introduced $d\Sigma_{\pt\mu}=-d\Sigma_\pt n_\mu$.
$\lambda^a$ and $\hcurrent_a^\mu$ denote sets of parameters, $\lambda^a(x)\equiv\{\beta^\mu(x), \nu(x)\}$, 
and of current operators, $\hcurrent_a^\mu(x)\equiv\{\hT^{\mu}_{~\nu}(x), \hJ^{\mu}(x)\}$, respectively.
Just as in the global case~(\ref{eq:GibbsDistribution}),
the Massieu-Planck functional $\Psi[\pt;\lambda]\equiv \ln\tr\exp(- \hK[\pt;\lambda])$
determines the normalization of the density operator $\hrho_{\text{LG}}$.
For constant parameters and $n_\mu=(-1,\bm{0})$, the local Gibbs distribution reproduces the global one~(\ref{eq:GibbsDistribution}).
We note that the definition here is coordinate invariant by construction.

The charge density operators on the hypersurface, $\hc_a(x)=\{\hp_\mu(x),\hn'(x) \}$, read $\hp_{\mu}(x)\equiv-n_\nu(x)\hT^\nu_{~\mu}(x)$ and $\hn'(x)\equiv-n_\nu(x)\hJ^\nu(x)$. 
Their expectation values, $\averageLG{\hc_a (x)}_\pt\equiv \tr\left[\hrho_{\text{LG}}[\pt; \lambda]\hc_a(x)\right]$,
are obtained from the variation of $\Psi[\pt;\lambda]$ with respect to $\lambda^a(x)$ on $\Sigma_\pt$,
\begin{equation}
\begin{split}
c_a(x)\equiv\averageLG{\hc_a(x)}_\pt=\frac{\delta}{\delta \lambda^a(x)}\Psi[\pt;\lambda]. \label{eq:Ca}
\end{split}
\end{equation}
The entropy is defined by
\begin{equation}
\begin{split}
S[\pt;c] &\equiv -\tr \hrho_{\text{LG}}[\pt; \lambda] \ln \hrho_{\text{LG}}[\pt; \lambda]\\
&=\averageLG{\hS[\pt;\lambda]}_\pt\\
&= -\int d\Sigma_\pt \lambda^a c_a+ \Psi[\pt;\lambda].
\end{split}
\end{equation}
The entropy is a functional of $c_a$, not $\lambda^a$, which
 can be confirmed by conducting the variation of $S$ with the fixed $\pt$,
\begin{equation}
\begin{split} 
\delta S &= \int d\Sigma_\pt\Bigl(- \delta\lambda^a c_a - \lambda^a \delta c_a
+ \frac{\delta \Psi[\pt;\lambda]}{\delta \lambda^a}\delta\lambda^a\Bigr)\\
&= -\int d\Sigma_\pt \lambda^a \delta c_a.
\end{split}
\end{equation}
The parameters are obtained as
\begin{equation}
\begin{split}
\lambda^a(x) =-\frac{\delta }{\delta c_a(x)}S[\pt;c]. \label{eq:delSdelC}
\end{split}
\end{equation}
For later purposes, we introduce $\psi^\mu$ such that  
\begin{equation}
\begin{split}
\Psi[\pt;\lambda] =\int d\Sigma_{\pt\mu} \psi^\mu=\int d\Sigma_{\pt} \psi,
\label{eq:pressure}
\end{split}
\end{equation}
where $\psi=-n_\mu\psi^\mu$, which satisfies
\begin{equation}
\begin{split}
d \psi = c_ad\lambda^a =p_\mu d\beta^\mu +n'd\nu, \label{eq:dpsi}
\end{split}
\end{equation} 
up to the covariant total derivative that does not contribute to $\delta\Psi$.
As will be seen in Sec.~\ref{sec:ZerothOrder}, in the leading order of derivative expansion, we can write $\psi^{\mu}$ as $\psi^\mu = \beta^\mu p(\beta,\nu)$ with the pressure $p$.
We note that there is an ambiguity in the definition of $\psi^\mu$ because $\Psi$ is invariant under the transformation $\psi^\mu\to \psi^\mu+g^\mu$ with
a function $g^\mu$ satisfying  $n_\mu g^\mu=0$.

Introducing the entropy current operator, 
\begin{equation}
\begin{split}
\hs^{\mu}\equiv - \lambda^a\hcurrent_a^\mu+\psi^\mu=
-\beta^\nu{\hT^{\mu}}_{~{\nu}} -\nu\hat{J}^{\mu} +\psi^\mu, \label{eq:entropyCurrent}
\end{split}
\end{equation}
the entropy reads
\begin{equation}
\begin{split}
S =\int d\Sigma_{\pt\mu} s^\mu=\int d\Sigma_\pt s\,,
\end{split}
\end{equation}
where $s^\mu\equiv \averageLG{\hs^{\mu}}_{\pt}$, and $s=-n_\mu s^\mu= -\lambda^ac_a+\psi=-\beta^\mu p_\mu -\nu n' +\psi$.
The entropy density $s$ satisfies the thermodynamic relation,
$ds = -\lambda^adc_a=-\beta^\mu dp_\mu -\nu dn'$, up to the covariant total derivative.
The divergence of the entropy current is
\begin{equation}
\begin{split}
\covDer_\mu \hs^\mu&=- (\covDer_\mu\lambda^a)\hcurrent_a^\mu+\covDer_\mu\psi^\mu,
\end{split}
\end{equation}
where we used the continuity equations $\covDer_\mu \hcurrent_a^\mu=0$.
In order to evaluate $\covDer_\mu \psi^\mu$, let us consider the derivative of $\Psi[\pt,\lambda]$ with respect to $\pt$, which reads
\begin{equation}
\begin{split}
\partial_\pt \Psi[\pt,\lambda]&=
-\averageLG{\partial_\pt \hK[\pt;\lambda] }_{\pt} \\
&= 
\Bigl\langle{ \partial_\pt\int d\Sigma_{\pt\mu}  \lambda^a\hcurrent_a^\mu\Bigr\rangle}^{\text{LG}}_\pt\\
&=\Bigl\langle{ \int d\Sigma_\pt N\covDer_\mu \bigl( \lambda^a\hcurrent_a^\mu\bigr)\Bigr\rangle}^{\text{LG}}_\pt\\
&= \int d\Sigma_\pt N (\covDer_\mu\lambda^a)\averageLG{\hcurrent_a^\mu}_\pt,
\label{eq:delPsi}
\end{split}
\end{equation}
where we again used the continuity equations. 
We also used  
\begin{equation}
\begin{split}
\partial_\pt \int d\Sigma_{\pt\mu}  f^\mu =  \int d\Sigma_\pt N \covDer_\mu f^\mu,  \label{eq:StokesTheorem}
\end{split}
\end{equation}
for an arbitrary smooth function $f^{\mu}(x)$ (see Appendix~\ref{sec:StokesTheorem}).

From Eq.~\eqref{eq:delPsi}, we obtain the divergence of $\psi^\mu$ as
\begin{equation}
\begin{split}
\covDer_\mu \psi^\mu =(\covDer_\mu\lambda^a)\averageLG{\hcurrent_a^\mu}_\pt.
\end{split}
\end{equation}
Then, the divergence of the entropy current operator reads
\begin{equation}
\begin{split}
\covDer_\mu \hs^\mu = -(\covDer_\mu \lambda^a)\delta \hcurrent^\mu_a
= -(\covDer_\mu \beta^\nu)\delta \hT^{\mu}_{~\nu}- (\covDer_\mu \nu)\delta \hJ^{\mu},
\label{eq:divS}
\end{split}
\end{equation}
where $\delta \hO\equiv\hO-\averageLG{\hO}_\pt$. The entropy production rate $\average{\covDer_\mu \hs^\mu}$ is in general nonzero.
When we decompose the expectation value of the current as $\average{\hcurrent^{\mu}_{a}} =\averageLG{\hcurrent^{\mu}_{a} }_\pt+ \average{\delta\hcurrent^{\mu}_{a}}$, $\averageLG{\hcurrent^{\mu}_{a} }_\pt$ can be identified as the nondissipative part because it does not contribute to the entropy production rate, while $ \average{\delta\hcurrent^{\mu}_{a}}$ can be identified as the dissipative part.

\subsection{Path-integral formulation of Massieu-Planck functional and thermal metric}\label{sec:pathIntegral}
In this subsection, we derive the path-integral formula for the Massieu-Planck functional $\Psi$. 
We show that the action has a form in the curved spacetime background, whose metric depends on
parameters $\beta^\mu$ and $\nu$.
We also show that the result is in accordance with those of recent studies, 
in which the Massieu-Planck functional is derived on the basis of symmetric and scaling properties~\cite{Banerjee:2012iz,Jensen:2012jh}.
Although we only consider a neutral scalar field here, the discussion covers 
the essential feature of the Massieu-Planck functional.

\begin{figure}[h]
\includegraphics[width=0.75\linewidth]{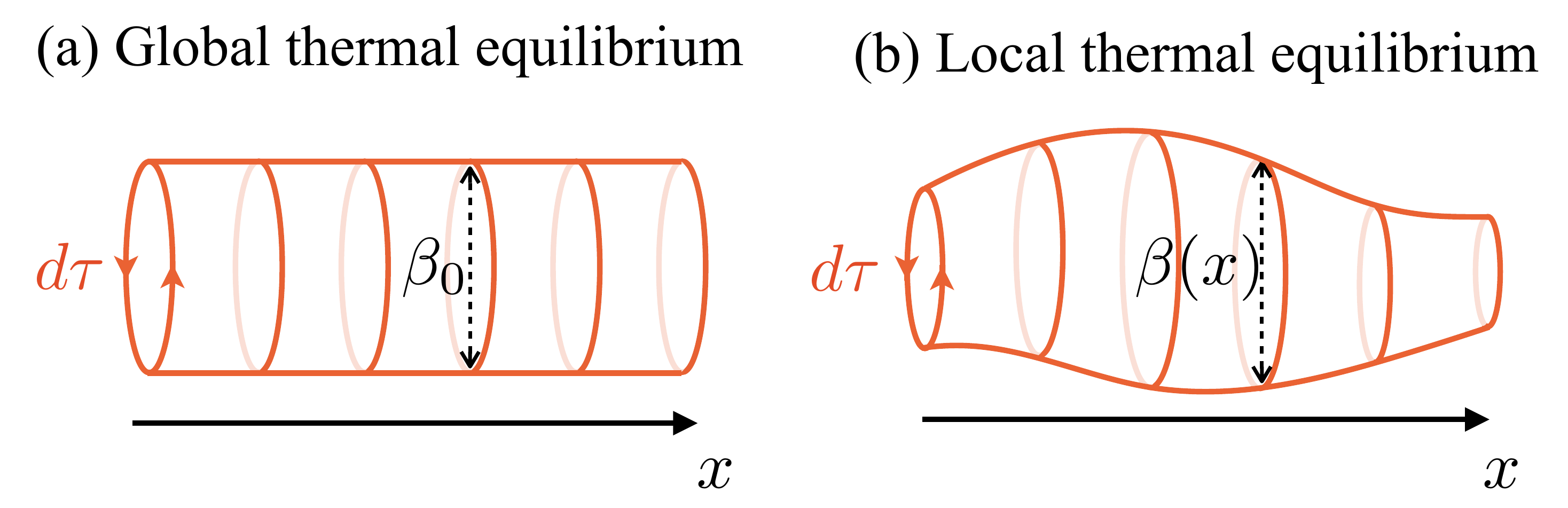}
\caption{
Comparison between the global thermal equilibrium~(a) and 
local thermal equilibrium states~(b). 
} 
\label{Fig:path-integral}
\end{figure}

In the coordinate system $(\pt,\bm{\px})$ with the ADM metric~\eqref{ADM}, the Lagrangian for a neutral scalar field $\phi$ reads
\begin{equation}
\begin{split}
\mathcal{L} = -\frac{g^{\ptc{\mu}\ptc{\nu}}}{2}\partial_\ptc{\mu}\phi \partial_\ptc{\nu}\phi  - V(\phi)
 = \frac{1}{2N^2}(\partial_\pt\phi -N^{\ptc{i}}\partial_\ptc{i}\phi)^2-
 \frac{\gamma^{\ptc{i}\ptc{j}}}{2}\partial_{\ptc{i}}\phi\partial_{\ptc{j}}\phi-V(\phi) ,
\end{split}
\end{equation}
where $V(\phi)$ denotes the potential term.
The canonical momentum $\pi(\bm{x})$ is $\pi(\bm{x}) \equiv -g^{\ptc{0}\ptc{\nu}}\partial_\ptc{\nu}\phi(\bm{\px})=N^{-2}(\partial_{\ptc{0}}\phi-N^{\ptc{i}}\partial_{\ptc{i}}\phi)$, which satisfies the canonical commutation relation, $[\hat{\phi}(\bm{\px}),\hpi(\bm{\px}')]=i\delta(\bm{\px}-\bm{\px}')/(N\sqrt{\gamma})$. 
We obtain the energy-momentum tensors as
\begin{align}
{\hT^{\ptc{0}}}_{~\ptc{0}} &= \hpi \partial_{\pt}\hat{\phi}-\hat{\mathcal{L}}=
\frac{N^{2}}{2}\hpi^2+N^{\ptc{i}} \hpi \partial_{\ptc{i}}\hat{\phi}+
\frac{\gamma^{\ptc{i}\ptc{j}}}{2}\partial_{\ptc{i}}\hat{\phi}\partial_{\ptc{j}}\hat{\phi}+\hat{V}(\phi), \\
{\hT^\ptc{0}}_{~\ptc{i}} &= \hpi \partial_\ptc{i} \hat{\phi}.
\end{align}
By using the standard technique of the path integral, we have  
\begin{equation}
\begin{split}
\tr e^{-\hK}&=\int d\phi\langle\phi|e^{- \hK}|\phi \rangle\\
&  = 
\int \mathcal{D}\phi \mathcal{D}\pi\exp\left(\int_0^{\beta_0} d\tau [i\int d^{d-1}\px N\sqrt{\gamma} \partial_\tau\phi(\tau,\bm{\px})\pi(\tau,\bm{\px}) -K]\right),
\label{eq:treK}
\end{split}
\end{equation}
where $K$ denotes the functional corresponding to the operator $\hK$.
After parametrizing $\beta^\ptc{\mu} = \beta_0 e^{\sigma}u^\ptc{\mu}$ and integrating Eq.~\eqref{eq:treK} with respect to $\pi$, 
we obtain the path-integral formula for the Massieu-Planck functional as
\begin{equation}
\begin{split}
\Psi[\pt; \lambda]
=  \ln\int \mathcal{D}\phi e^{+S[\phi,\lambda]},\\
\end{split}
\end{equation}
with
\begin{equation}
\begin{split}
S[\phi,\lambda] 
&=\int_0^{\beta_{0}} d\tau\int d^{d-1}\px\sqrt{\gamma}\tilde{N} \Bigl[
\frac{1}{2\tilde{N}^2}\Bigl(i\partial_\tau \phi -\tilde{N}^{\ptc{i}}\partial_{\ptc{i}}\phi \Bigr)^2
-\Bigl(
\frac{\gamma^{\ptc{i}\ptc{j}}}{2}\partial_{\ptc{i}}\phi\partial_{\ptc{j}}\phi+V(\phi)\Bigr) 
\Bigr]\\
&\equiv\int_0^{\beta_{0}} d\tau \int d^{d-1}\px \sqrt{-\tilde{g}}\tilde{\mathcal{L}}(\phi, \partial_\ptc{\rho}\phi; \tilde{g}_{\ptc{\mu}\ptc{\nu}}),
\end{split} 
\end{equation}
where $\tilde{N}\equiv Nu^{\ptc{0}} e^{\dilatation}=-n_\mu\beta^\mu/\beta_0$, $\tilde{N}^{\ptc{i}}\equiv \gamma^{\ptc{i}\ptc{j}}e^{\sigma}u_\ptc{j}$.
We define the thermal metric $\tilde{g}_{\ptc{\mu}\ptc{\nu}}$ and its inverse $\tilde{g}^{\ptc{\mu}\ptc{\nu}}$ as
\begin{equation}
\begin{split}
\tilde{g}_{\ptc{\mu}\ptc{\nu}} 
=
\begin{pmatrix}
-\tilde{N}^2 +\tilde{N}_\ptc{i}\tilde{N}^\ptc{i} & \tilde{N}_\ptc{j}\\
\tilde{N}_{\ptc{i}}&  \gamma_{\ptc{i}\ptc{j}}
\end{pmatrix},\qquad
\tilde{g}^{\ptc{\mu}\ptc{\nu}}=
\begin{pmatrix}
-\tilde{N}^{-2}& \tilde{N}^{-2}\tilde{N}^{\ptc{j}}\\
\tilde{N}^{-2} \tilde{N}^{\ptc{i}}&  \gamma^{\ptc{i}\ptc{j}}-\tilde{N}^{-2}\tilde{N}^{\ptc{i}}\tilde{N}^{\ptc{j}}
\end{pmatrix}.
\end{split} \label{eq:thermalMetric}
\end{equation}
Here, $\tilde{N}_\ptc{i} \equiv \gamma_{\ptc{i}\ptc{j}}\tilde{N}^\ptc{j} = e^\sigma u_\ptc{i} $. 
This metric again has the form of the ADM metric:
\begin{equation}
 \begin{split}
  ds^2 = -(\tilde{N}d\tilde{t})^2 +\gamma_{\ptc{i}\ptc{j}}(\tilde{N}^\ptc{i}d\tilde{t}+d\px^\ptc{i})(\tilde{N}^\ptc{j}d\tilde{t}+d\px^\ptc{j}),
 \end{split}
\end{equation}
with $d\tilde{t}=-id{\tau}$. 
In Fig.~\ref{Fig:path-integral}, we show a schematic figure of a locally thermalized state by comparing it with that of the globally thermalized one.
While the (uniform) thermal field theory is formulated under the flat spacetime as shown in Fig.~\ref{Fig:path-integral}(a), 
the locally thermalized field theory can be formulated under a curved spacetime background. 
The metric is determined by the thermodynamic parameters such as 
the temperature, and the fluid four-velocity as in Eq.~\eqref{eq:thermalMetric}, 
and thus the imaginary-time radius manifestly depends on the spacetime as shown in Fig.~\ref{Fig:path-integral}(b).
The line element,
$ds^2$ is not real because $d\tilde{t}$ is imaginary, so that the action $S[\phi,\lambda]$ is in general complex, 
which causes the sign problem in lattice simulations.
This expression of the thermal metric does not explicitly depend on the
choice of the original shift vector $N^{\ptc{i}}$.

The thermal metric is invariant under the imaginary-time translation, since
the parameters $\lambda^a$ do not depend on the imaginary time $\tau$, $\tilde{t}=-i\tau$.
Furthermore, we also have local symmetry by the redefinition of the imaginary time.
In order to demonstrate this symmetry, we rewrite the thermal metric from the ADM form to the Kaluza-Klein one as
\begin{equation}
\begin{split}
ds^2 
&= -e^{2\sigma}(d\tilde{t}+a_{\ptc{i}}d\px^{\ptc{i}})^2 +\gamma'_{\ptc{i}\ptc{j}}d\px^\ptc{i} d\px^\ptc{j} , 
\end{split}
\end{equation}
where $a_{\ptc{i}}\equiv  -e^{-\sigma}u_\ptc{i} $, ${\gamma'}_{\ptc{i}\ptc{j}}\equiv\gamma_{\ptc{i}\ptc{j}}+u_\ptc{i}u_\ptc{j}$, and
we used $\tilde{g}_{\ptc{0}\ptc{0}} = -\tilde{N}^2 +\tilde{N}_\ptc{i}\tilde{N}^\ptc{i} =-e^{2\sigma}$.
In this parametrization, the square root of determinant of metric becomes $\sqrt{-\tilde{g}}= \tilde{N}\sqrt{\gamma}=e^{\sigma}\sqrt{\gamma'}$.
This parametrization of the Massieu-Planck functional was discussed in Ref.~\cite{Banerjee:2012iz}.
Following Ref.~\cite{Banerjee:2012iz},
we can easily see that this metric is invariant under the local transformation (the Kaluza-Klein gauge transformation),
\begin{equation}
 \begin{cases}
  \tilde{t} \to \tilde{t} + \chi(\bm{\px}), \\
  \bm{\px} \to \bm{\px}, \\
  a_\ptc{i}(\bm{\px}) \to a_\ptc{i}(\bm{\px}) - \partial_\ptc{i} \chi(\bm{\px}), 
 \end{cases}
\end{equation}
where $\chi(\bm{\px})$ is an arbitrary function of the spatial coordinates.
We note that $\gamma_{\ptc{i}\ptc{j}}$ nonlinearly transforms under this transformation
since $\gamma'_{\ptc{i}\ptc{j}}$ does not change, so that $\gamma$ is not gauge invariant.
This symmetry enables us to restrict possible terms that appear in the Massieu-Planck functional~\cite{Banerjee:2012iz}.
For example, $a_\ptc{i}$ appears in the Massieu-Planck functional only through the gauge invariant combination such as
the field strength, $f_{\ptc{i}\ptc{j}} \equiv \partial_\ptc{i} a_\ptc{j} - \partial_\ptc{j} a_\ptc{i}$.

In addition to the above symmetry associated with the imaginary time translation, 
the Massieu-Planck functional has the $(d-1)$-dimensional spatial diffeomorphism, $\bm{\px}\to \bm{\px}'(\bm{\px})$.
This spatial diffeomorphism invariance also restricts possible terms that could appear in the Massieu-Planck functional. 
For example, $\gamma'$ appears only in combination with $d^{d-1}\px$, i.e., $d^{d-1} \px \sqrt{\gamma'}=d\Sigma_\pt N e^{-\sigma}$.
In Sec.~\ref{sec:DerivativeExpansion}, we will write down the possible form of the Massieu-Planck functional
within the derivative expansion using these symmetric properties.

Although we only consider the neutral scalar field, 
the extension to a system with finite chemical potential is straightforward: We may replace the partial derivative $\partial_\tau$ with the covariant one,
$D_\tau \equiv (\partial_\tau - e^\sigma \mu)$, 
in which the additional term $e^\sigma \mu=\nu/\beta_0$ is Kaluza-Klein gauge invariant.
Therefore, the symmetric properties of the thermal metric,
which are discussed in this subsection, also hold for systems with finite chemical potential.

\section{Time evolution} \label{sec:TimeEvolution}
In the previous section, we considered the local thermodynamics on the hypersurface.
Here, we discuss the time evolution of the expectation values of local operators.
In a quantum field theory, the expectation value of a local operator is given by
\begin{equation}
\begin{split}
\average{\hO(x)}= \tr \hrho_0 \hO(x),
\end{split}
\end{equation}
where $\hrho_0$ is the density operator at initial time. In particular, we consider the time evolution of hydrodynamic variables $c_a(x)$.
If the constitutive relation is obtained, i.e., 
if $\average{\hcurrent_a^\mu}$ is expressed as a functional of $c_a$ or $\lambda^a$, its time-evolution equation (hydrodynamic equation) is given by the continuity equation $\covDer_\mu \average{\hcurrent_a^\mu}=0$.
To obtain the constitutive relation, it is useful to decompose $\average{\hcurrent_a^\mu}$ into nondissipative and dissipative parts,
$\average{\hcurrent^{\mu}_{a}} =\averageLG{\hcurrent^{\mu}_{a} }_\pt+ \average{\delta\hcurrent^{\mu}_{a}}$. 
The nondissipative part $\averageLG{\hcurrent^{\mu}_{a} }_\pt$ is obviously a functional of $\lambda^a(x)$ and does not contain the information of the past state.
On the other hand, we need the information of the past to evaluate $\average{\delta\hcurrent_a^\mu}$.
The purpose of this section is to derive the self-consistent equation 
to determine $\average{\delta\hcurrent_a^\mu}$.

At a very early stage of time evolution, the system will be far from equilibrium in a state that cannot be characterized by only thermodynamic or hydrodynamic variables.
In this stage, microscopic degrees of freedom play an important role to determine the time evolution of the system.
In contrast, at later times, we expect the system to be characterized 
by the thermodynamic variables whose time evolution is governed by the hydrodynamic equations.
In this paper, we assume that at the time $\pt_0$, the distribution function is given by a local Gibbs one, $\hrho_0\equiv \hrho_{\text{LG}}[\pt_0;\lambda]$,
although, in general, this is not exact but only approximate. 
As we will see below, once we assume this initial condition,
the time-evolution equation can be rewritten as a compact form.

In order to evaluate the expectation value of $\delta \hcurrent^\mu_a(x)$ at the point $x^\mu\in\Sigma_\pt$ for $\pt>\pt_0$,
we decompose the density operator into the local Gibbs distribution on $\Sigma_\pt$ and the other:
\begin{equation}
\begin{split}
\hrho(\pt_0)=\exp\bigl(-{\hS[\pt_0;\lambda]}\bigr)
=\exp\bigl(-\hS[\pt;\lambda]+\hSigma[\pt,\pt_0;\lambda]\bigr),
\end{split}
\end{equation}
where $\hSigma[\pt,\pt_0;\lambda]\equiv\hS[\pt;\lambda]-\hS[\pt_0;\lambda]$.
$\hSigma[\pt,\pt_0;\lambda]$ can be expressed by the divergence of the entropy current operator as
\begin{equation}
\begin{split}
\hSigma[\pt,\pt_0;\lambda]
 &=\int_{\pt_0}^{\pt} d\ptc{s} \partial_{\ptc{s}}
\int d\Sigma_{\ptc{s}\mu}\hs^\mu =\int_{\pt_0}^\pt d\ptc{s} \int d\Sigma_\ptc{s} N\covDer_\mu\hs^{\mu}.
\label{eq:entropy1}
\end{split}
\end{equation}
In the last line, we used Eq.~\eqref{eq:StokesTheorem}.
The explicit form of $\covDer_\mu\hs^{\mu}$ is given in Eq.~\eqref{eq:divS}.

We will treat $\hSigma[\pt,\pt_0;\lambda]$ as the perturbation term in the derivative expansion 
because $\covDer_\mu\hs^{\mu}$ is proportional to the derivatives of the parameters, $\covDer_\mu\lambda^a$.
In order to expand $\hrho(\pt_0)$ with respect to $\hSigma[\pt,\pt_0;\lambda]$, we decompose the density operator as
\begin{equation}
\begin{split}
\hrho(\pt_0)= \hrho_{\text{LG}}(\pt) \hU(\pt,\pt_0),
\end{split}
\end{equation}
where $\hU(\pt,\pt_0)$ is defined as
\begin{equation}
\begin{split}
\hU(\pt,\pt_0)\equiv T_\tau e^{\int_0^{1} d\tau \hSigma_\tau[\pt,\pt_0;\lambda]},
\end{split}
\end{equation}
with $\hSigma_\tau[\pt,\pt_0;\lambda] \equiv e^{\tau \hK[\pt;\lambda]}\hSigma[\pt,\pt_0;\lambda] e^{-\tau \hK[\pt;\lambda]}$.
Here, $T_\tau$ denotes $\tau$ ordering.
The expectation value of an operator $\hO(x)$ on $\Sigma_\pt$ is given by  
\begin{equation}
\begin{split}
\langle\hO(x) \rangle= \averageLG{\hU  \hO(x) }_{\pt},
\label{eq:expectationValue}
\end{split}
\end{equation}
where $\averageLG{\hO(x)}_{\pt}\equiv \tr \hrho_{\text{LG}}[\pt; \lambda]\hO(x)$.
If one takes $\hO=\hU^{-1}$, Eq.~\eqref{eq:expectationValue} gives an identity corresponding to an integral  fluctuation theorem, $\average{\hU^{-1}}=1$~\cite{sasa:2013}.

Since Eq.~\eqref{eq:expectationValue} is the identity, it holds for any parameters $\lambda^a$.
We need a condition to fix $\lambda^a$. Here we impose $\average{\hc_a(x)}=\averageLG{\hc_a(x)}_\pt$~\cite{Zubarev:1979}; they are explicitly
\begin{align}
n_\mu(x) \average{\hT^{\mu}_{~{\nu}}(x)} &= n_\mu(x)  \averageLG{\hT^{\mu}_{~\nu}(x)}_{\pt}, \label{eq:condition1}\\
n_\mu(x) \average{ \hJ^{\mu}(x)} &= n_\mu(x)  \averageLG{ \hJ^{\mu}(x) }_{\pt}. \label{eq:condition2}
\end{align}
The parameters are determined by the entropy functional through Eq.~\eqref{eq:delSdelC}.
Equations~\eqref{eq:condition1} and \eqref{eq:condition2} mean that the dissipative parts $\average{\delta\hcurrent^{\mu}_{a}}$ are orthogonal to $n_\mu(x) $, i.e., $n_\mu\average{\delta\hcurrent^{\mu}_{a}}=-\average{\delta\hc_a} =0$.

In order to consider the time evolution, we use the spatial projection operator introduced in Sec.~\ref{sec:Geometric},
\begin{equation}
\begin{split}
P^\mu_\nu \equiv \delta^\mu_\nu+ v^\mu n_\nu
\quad
{\rm with}
\quad
v^\mu n_\mu=-1\,,
\quad
P^\mu_\nu v^\nu=0\,,
\quad
P^\mu_\nu n_\mu=0\,.
\end{split}
\end{equation}
Then, the derivative is written as
\begin{equation}
\begin{split}
\covDer_\mu = (-v^\nu n_\mu+P^\nu_\mu)\covDer_\nu = -\frac{n_\mu}{N} \delt +\covDer_{\perp\mu}, \label{eq:derivativeDecomposition}
\end{split}
\end{equation}
where $\delt= Nv^\mu\covDer_\mu$ and $\covDer_{\perp\mu} \equiv  P_{\mu}^{~\nu}\covDer_\nu$.
By using this projection operator, $\hSigma[\pt,\pt_0;\lambda]$ reads as
\begin{equation}
\begin{split}
\hSigma[\pt,\pt_0;\lambda] &=-\int_{\pt_0}^\pt d\ptc{s} \int d\Sigma_\ptc{s} 
\Bigl[
(\covDer_{\ptc{s}} \lambda^a)\delta \hc_{a}+N(\covDer_{\perp\mu} \lambda^a)\delta \hcurrent^{\mu}_{a}
\Bigr].
 \end{split}
\end{equation}
We would like to eliminate the time derivative of parameters $\covDer_{\ptc{s}} \lambda^a$ from $\hSigma[\pt,\pt_0;\lambda]$, which can be performed by using
the continuity equation, $\covDer_\mu\average{\hcurrent^{\mu}_{a}}=\covDer_\mu \averageLG{\hcurrent^{\mu}_{a} }_\pt+ \covDer_\mu \average{\delta\hcurrent^{\mu}_{a}}=0$.
Since $\hS[\pt;\lambda]$ does not depend on $\px^{i}$, i.e., $\covDer_{\perp\mu}\hS[\pt;\lambda]=0$,
$\covDer_\mu \hS[\pt;\lambda]= -(n_{\mu}/N)\partial_{\pt}\hS [\pt;\lambda]$,
we can write
the divergence of  $\averageLG{\hcurrent^{\mu}_{a} }_\pt$ as
\begin{equation}
\begin{split}
\covDer_\mu \averageLG{\hcurrent^{\mu}_{a}(x) }_\pt 
&= \tr \Bigl[\frac{1}{N(x)} \bigl(\partial_{\pt} e^{-\hS[\pt;\lambda]} \bigr)\hc_{a}(x)\Bigr]\\
&= \frac{-1}{N(x)} \int d\Sigma'_\pt N(x')\int_0^1 d\tau\averageLG{e^{\tau\hK[\pt;\lambda]}\covDer_\mu \hs^\mu(x') e^{-\tau\hK[\pt;\lambda]} \hc_{a}(x)}_\pt\\
&= \frac{1}{N(x)}\int d\Sigma'_\pt N(x') (\covDer_\nu\lambda^b(x')) \bm{(} \delta\hc_a(x),{\delta\hcurrent^{\nu}_b}(x')\bm{)}_\pt, \label{eq:delJ}
\end{split}
\end{equation}
where  $\bm{(}\hat{A},\hat{B}\bm{)}_{\pt}$ is the local Gibbs version of the Kubo-Mori-Bogoliubov inner product,
\begin{equation}
\begin{split}
\bm{(}\hat{A},\hat{B}\bm{)}_{\pt}\equiv 
\int_0^1 d\tau \averageLG{e^{\hK\tau} \hat{A}e^{-\hK\tau}\hat{B}^\dag}_{\pt}, 
\label{eq:innerProd}
\end{split}
\end{equation}
which satisfies linearity $\bm{(}a\hat{A}+b\hat{B},\hat{C}\bm{)}_{\pt}=a\bm{(}\hat{A},\hat{C}\bm{)}_{\pt}+b\bm{(}\hat{B},\hat{C}\bm{)}_{\pt}$, Hermite symmetry $\bm{(}\hat{A},\hat{B}\bm{)}_{\pt}^*=\bm{(}\hat{B},\hat{A}\bm{)}_{\pt}$,
and positivity $\bm{(}\hat{A},\hat{A}\bm{)}_{\pt}\geq0$; $\bm{(}\hat{A},\hat{A}\bm{)}_{\pt}=0\Rightarrow\hat{A}=0$.
We used $ \bm{(} \averageLG{\hc_a(x)}_\pt,{\delta\hcurrent^{\nu}_b}(x')\bm{)}_\pt= \averageLG{\hc_a(x)}_\pt\averageLG{\delta\hcurrent^{\nu}_b(x')}_\pt=0$ 
to obtain the last line in Eq.~\eqref{eq:delJ}.
Using Eq.~\eqref{eq:derivativeDecomposition}, we find that $\covDer_\mu\average{\hcurrent^{\mu}_{a}}=0$ leads to 
\begin{equation}
\begin{split}
&\int d\Sigma'_\pt
\bm{(} \delta\hc_a(x),\delta\hc_b(x')\bm{)}_\pt \delt\lambda^b(x')\\
&\quad+
\int d\Sigma'_\pt  
\bm{(} \delta\hc_a(x),{\delta\hcurrent^{\nu}_{b}}(x')\bm{)}_\pt N(x')\covDer_{\perp\nu}\lambda^b(x')
+N(x)\covDer_\mu \average{\delta\hcurrent^{\mu}_{a}(x) }
=0. \label{eq:identity1}
\end{split}
\end{equation}
Multiplying Eq.~\eqref{eq:identity1} by the inverse of $\bm{(} \delta\hc_a(x),\delta\hc_b(x')\bm{)}_\pt$, and integrating it with respect to the coordinates on the hypersurface,
we obtain
\begin{equation}
\begin{split}
 \delt\lambda^a(x)&= -\int d\Sigma'_\pt \int d\Sigma''_\pt  
\bm{(} \delta\hc_a(x),\delta\hc_b(x')\bm{)}_\pt^{-1} \bm{(} \delta\hc_b(x'),{\delta\hcurrent^{\nu}_{c}}(x'')\bm{)}_\pt N(x'')\covDer_{\perp\nu}\lambda^c(x'')\\
&\quad-
 \int d\Sigma'_\pt  \bm{(} \delta\hc_a(x),\delta\hc_b(x')\bm{)}_\pt^{-1} N(x')\covDer_\mu \average{\delta\hcurrent^{\mu}_{b}(x') }. \label{eq:timeEvolution}
\end{split}
\end{equation}
Let us eliminate $\covDer_{\ptc{s}}\lambda^a$ in $\Sigma[\pt,\pt_0;\lambda]$.
For this purpose, it is convenient to introduce a projection operator $\hProj$
onto $\delta\hc_a$,
\begin{equation}
\begin{split}
\hProj\hO =  \int d\Sigma_\pt  \int d\Sigma'_\pt \delta\hc_a(x) \bm{(}\delta\hc_a(x),\delta\hc_b(x') \bm{)}_\pt^{-1} \bm{(}\delta\hc_b(x'),\hO \bm{)}_\pt. \label{eq:MoriProjection}
\end{split}
\end{equation}
This is the relativistic version of the projection operator used in Refs.~\cite{PhysRevA.8.2048,Esposito:1994rt}. 
At thermal equilibrium, it reduces to the Mori projection operator~\cite{Mori}.
We have
\begin{align}
 \bm{(}\delta\hc_b(x'),\hO \bm{)}_\pt &= \frac{\delta}{\delta\lambda^b(x')} \averageLG{\hO}_\pt,
 \label{eq:projection1}\\
 \bm{(}\delta\hc_a(x),\delta\hc_b(x') \bm{)}_\pt^{-1}&= \frac{\delta \lambda^b(x')}{\delta c_a(x)}.
\label{eq:projection2}
\end{align}
Using Eqs.~\eqref{eq:projection1} and~\eqref{eq:projection2} and the chain rule, we can rewrite Eq.~\eqref{eq:MoriProjection} as
\begin{equation}
\begin{split}
\hProj\hO =  \int d\Sigma_\pt  \int d\Sigma'_\pt \delta\hc_a(x)\frac{\delta \lambda^b(x')}{\delta c_a(x)} \frac{\delta}{\delta\lambda^b(x')} \averageLG{\hO}_\pt
=  \int d\Sigma_\pt  \delta\hc_a(x)\frac{\delta }{\delta c_a(x)}  \averageLG{\hO}_\pt.
\end{split}
\end{equation}

Now, by using $\hProj$, 
we can eliminate $\delt\lambda^a$ from $\hSigma[\pt,\pt_0;\lambda] $, and we obtain
\begin{equation}
\begin{split}
\hSigma[\pt,\pt_0;\lambda] &=-\int_{\pt_0}^\pt d\ptc{s} \int d\Sigma_\ptc{s}N\Bigl[ (\covDer_{\perp\mu}\lambda^a)(1-\hProj) \delta\hcurrent_a^\mu
- \delta\hlambda^a\covDer_\mu \average{\delta\hcurrent^{\mu}_{a} }\Bigr]
\\
&=-\int_{\pt_0}^\pt d\ptc{s} \int d\Sigma_\ptc{s} N\Bigl[
(\covDer_{\perp\mu}\beta_\nu ) {\delQ \hT^{\mu\nu}} + (\covDer_{\perp\mu}\nu) {\delQ\hJ^{\mu}}-\delta\hlambda^a\covDer_\mu \average{\delQ\hcurrent^{\mu}_{a} }\Bigr].
\end{split}
\end{equation}
Here we introduced $\delQ\hO\equiv   (1-\hProj)\delta\hO$, which enables us to remove the hydrodynamic modes from
$\delta\hO$. 
In the second line, we replaced $ \average{\delta\hcurrent^{\mu}_{a} }$ by $ \average{\delQ\hcurrent^{\mu}_{a} }$ because 
 the expectation value of the projected operator vanishes, $\average{\hProj\hO}=0$.
We also defined
\begin{equation}
\begin{split}
\delta\hlambda^a(x) \equiv  \int d\Sigma'_\pt \delta\hc_b(x') \frac{\delta \lambda^a(x)}{\delta c_b(x')}.
\end{split}
\end{equation}

For later use, we perform the tensor decomposition for $\delQ \hT^{\mu\nu}$.
Since $n_\mu\delQ \hT^{\mu\nu}=0$ and $n_\nu\delQ \hT^{\mu\nu}=0$,
we can decompose $\delQ \hT^{\mu\nu}$ as $\delQ \hT^{\mu\nu} =  h^{\mu\nu} \delQ\hp + \delQ \hpi^{\mu\nu}$,
where 
\begin{align}
\delQ\hp&\equiv \frac{1}{d-1}h_{\rho\sigma}\delQ \hT^{\rho\sigma},\\
\delQ\hpi^{\mu\nu}&\equiv P^{\mu}_{\rho} P^{\nu}_{\sigma} \delQ \hT^{\rho\sigma} - \frac{h^{\mu\nu}}{d-1}h_{\rho\sigma}\delQ \hT^{\rho\sigma}.
\end{align}
Here we introduced $h^{\mu\nu}\equiv P^{\mu}_{\rho} P^{\nu}_{\sigma}g^{\rho\sigma}$ 
and $h_{\mu\nu}$ that satisfy $h^{\mu\rho}h_{\rho\nu}=P^\mu_\nu$.

As a result, $\hSigma[\pt,\pt_0;\lambda] $ reads
\begin{equation}
\begin{split}
\hSigma[\pt,\pt_0;\lambda] 
&=-\int_{\pt_0}^\pt d\ptc{s} \int d\Sigma_\ptc{s} N\Bigl[
(h^{\mu\nu}\covDer_{\mu}\beta_\nu) {\delQ \hp} +(\covDer_{\langle\mu} \beta_{\nu\rangle}) \delQ \hpi^{\mu\nu}+ (\covDer_{\perp\mu}\nu) {\delQ\hJ^{\mu}}-\delta\hlambda^a\covDer_\mu \average{\delQ\hcurrent^{\mu}_{a}}\Bigr],
\label{eq:EntropyProduction}
\end{split}
\end{equation}
where 
\begin{align}
\covDer_{\langle\mu}\beta_{\nu\rangle} &\equiv \frac{P_{\mu}^\rho P_\nu^{\sigma}}{2}(\covDer_{\rho}\beta_{\sigma}+\covDer_{\sigma}\beta_{\rho})     -\frac{h_{\mu\nu}}{d-1}h^{\rho\sigma}\covDer_{\rho}\beta_\sigma.
\end{align}
We note that $\covDer_\mu \average{{\delQ}\hcurrent^{\mu}_{a}}$ does not contain the explicit time derivative of the parameters because
$\covDer_\mu \average{{\delQ}\hcurrent^{\mu}_{a}}=(-N^{-1}n_\mu \delt+\covDer_{\perp\mu}) \average{{\delQ}\hcurrent^{\mu}_{a}} =\bigl(N^{-1}(\delt n_\mu) +\covDer_{\perp\mu}\bigr) \average{{\delQ}\hcurrent^{\mu}_{a}}$,
where we used $n_\mu\delt \average{{\delQ}\hcurrent^{\mu}_{a}}=-(\delt n_\mu) \average{{\delQ}\hcurrent^{\mu}_{a}}$.

Since $\average{\delta\hcurrent^{\mu}_{b}(x)} =\average{{\delQ}\hcurrent^{\mu}_{b}(x) }$, 
 our goal is now to solve 
\begin{equation}
\begin{split}
\average{{\delQ}\hcurrent^{\mu}_{b}(x) }= \averageLG{T_\tau e^{\int_0^{1} d\tau \hSigma_\tau[\pt,\pt_0;\lambda]}{\delQ}\hcurrent^{\mu}_{b}(x) }_\pt.
\label{eq:deltaJ}
\end{split}
\end{equation}
$\hSigma_\tau[\pt,\pt_0;\lambda]$ contains $\average{{\delQ}\hcurrent^{\mu}_{b}(x)}$ as in Eq.~\eqref{eq:EntropyProduction}, 
so that Eq.~\eqref{eq:deltaJ} becomes a self-consistent equation.
As we discuss in the next section, $\average{{\delQ}\hcurrent^{\mu}_{b}(x) }$ can be evaluated order by order in the derivative expansion with respect to the parameters.

\section{Derivative expansion and hydrodynamic equations} \label{sec:DerivativeExpansion}
In this section we perform the derivative expansion to derive relativistic hydrodynamic equations order by order.
We also discuss the frame choice, which originates from an ambiguity in the definition of the fluid four-velocity.

\subsection{Derivative expansion} \label{sec:TwoExpansion}
The expectation value of $\hcurrent_a^\mu (x)$ consists of the nondissipative and dissipative parts, $\average{\hcurrent_a^\mu (x)}=\averageLG{\hcurrent_a^\mu(x)}_\pt+\average{\delQ \hcurrent_a^\mu(x)}$. 
As will be shown in the following,
the nondissipative part $\averageLG{\hcurrent_a^\mu(x)}_\pt$ is obtained by differentiating the Massieu-Planck functional $\Psi$ with respect to $\pt$. 
The Massieu-Planck functional  can be expanded as
\begin{align}
\Psi[\lambda] &= \sum_{n=0}^\infty \Psi^{(n)}[\lambda],  
\end{align}
where $n$ denotes the order of spatial derivative ${O}(\covDer_\perp^{n})$~\footnote{On curved space, curvatures may appear in higher-derivative terms. 
For example, we identify the spatial curvature as the second-order derivative, because it is given by a commutator of the spatial covariant derivatives.}.
As was discussed in Sec.~\ref{sec:pathIntegral},
$\Psi[\lambda]$ and therefore $\Psi^{(n)}[\lambda]$
enjoy thermal Kaluza-Klein symmetry and spatial diffeomorphism invariance.
For parity symmetric theories, $\Psi^{(1)}[\lambda]$ vanishes
because we cannot construct a scalar with one spatial derivative
such that it is invariant under the above symmetries.
On the other hand,  the higher-order terms  are not forbidden by parity symmetry. 
The second-or higher-order hydrodynamics can contain nondissipative terms coming from them. 
The general expansion of the nondissipative parts based on these symmetries was discussed in Ref.~\cite{Banerjee:2012iz}.

The dissipative part $ \average{\delQ\hcurrent_a^\mu(x)}$ can be expanded as
\begin{equation}
\begin{split}
 \average{\delQ\hcurrent_a^\mu(x)} &=\sum_{m,n=0}^\infty\average{\delQ \hcurrent^{\mu}_a (x)}_{(m,n)},
\end{split}
\end{equation}
where the term labeled by $(m,n)$
contains $m$ temporal derivatives, $\delt$,
and $n$ spatial derivatives, $\covDer_{\perp}$.
In order to evaluate $\average{\delQ \hcurrent^{\mu}_a (x)}_{(n,m)}$, we expand the dissipative part $\average{\delQ\hcurrent_a^\mu (x)}$ as 
\begin{equation}
\begin{split}
\average{\delQ\hcurrent_a^\mu(x)} &= \averageLG{T_\tau  e^{\int_0^{1}d\tau\hSigma_\tau(\pt,\pt_0)}\delQ\hcurrent_a^\mu (x)}_{\pt} \\
 &= \averageLG{\delQ\hcurrent_a^\mu (x)}_{\pt} + \int_0^{1}d\tau\averageLG{T_\tau {\hSigma_\tau(\pt,\pt_0)}\delQ\hcurrent_a^\mu (x)}_{\pt}\\
&\qquad\qquad\qquad\,\; + \frac{1}{2}\int_0^{1}d\tau \int_0^{1}d\tau' \averageLG{T_\tau \hSigma_\tau(\pt,\pt_0)\hSigma_{\tau'}(\pt,\pt_0)\delQ\hcurrent_a^\mu (x)}_{\pt} +\cdots .
\end{split}
\label{eq:Expansion}
\end{equation}
Here $\averageLG{\delQ\hcurrent_a^\mu (x)}_{\pt}$ vanishes by definition.
Since $\hSigma_\tau(\pt,\pt_0)$ contains the derivative of the parameters, $\covDer_\perp\lambda^a$, $\hSigma_\tau(\pt,\pt_0)$ is identified as of order $\covDer_\perp$.
We note that $\hSigma_\tau(\pt,\pt_0)$ does not contain the temporal derivative of the parameters, $\delt \lambda$.
This fact implies that the derivative expansion starts from $\average{\delQ\hcurrent^{\mu}_a(x)}_{(0,1)}$; i.e., $\average{\delQ \hcurrent^{\mu}_a(x)}_{(l,0)}$ for $l\geq0$ vanishes.
If one considers the $n$th order of $ \average{\delQ\hcurrent_a^\mu(x)}$, one may expand Eq.~\eqref{eq:Expansion} up to the $n$th order of $\hSigma_\tau(\pt,\pt_0)$.
All correlation functions with lower orders of $\hSigma_\tau(\pt,\pt_0)$ contribute to the $n$th order of $\average{\delQ\hcurrent_a^\mu(x)}$. 
For example, in addition to the third term in the second line of Eq.~\eqref{eq:Expansion},
the second term contributes to $\average{\delQ\hcurrent^{\mu}_a(x)}_{(0,2)}$ through the derivative expansion of the correlation function $\averageLG{T_\tau {\hSigma_\tau(\pt,\pt_0)}\delQ\hcurrent_a^\mu (x)}_{\pt}$. 
In the following, we restrict ourselves to the zeroth and first-order hydrodynamic equations with parity symmetry.
 
\subsubsection{Zeroth order: Perfect fluid}\label{sec:ZerothOrder}
Let us consider the leading order of $\average{\hcurrent^\mu_a}$ in the derivative expansion.
We show that the energy-momentum tensor and the current have the form of a perfect fluid.
In Sec.~\ref{sec:pathIntegral}, we discussed that the Massieu-Planck functional is obtained from the path integral in curved spacetime, 
whose metric $\tilde{g}_{\ptc{\mu}\ptc{\nu}}$ is invariant under the thermal Kaluza-Klein transformation.
Thanks to the Kaluza-Klein gauge symmetry, $\Psi^{(0)}[\lambda]$ does not contain $a_\ptc{i}$. 
Furthermore, the spatial diffeomorphism invariance restricts the $\gamma$ dependence of $\Psi^{(0)}[\lambda]$ to the form proportional to  $d^{d-1}\px\sqrt{\gamma'}$,
while it does not restrict  the $\sigma$ dependence of $\Psi^{(0)}[\lambda]$.
Then, we factorize $\Psi^{(0)}[\lambda]$ as~\cite{Banerjee:2012iz} 
\begin{equation}
  \begin{split}
    \Psi^{(0)}[\lambda] &=\int_0^{\beta_0} d\tau \int d^{d-1} \px\; e^\sigma \sqrt{\gamma'} p(\beta,\nu), \\
     &= \int d\Sigma_\pt\;  \beta' p(\beta,\nu),
  \end{split}
\end{equation}
where $\beta'\equiv -n_\mu\beta^\mu$, $\beta=\beta_0e^{\sigma}$, and $p(\beta,\mu)$ is the pressure of the perfect fluid as explicitly shown later.
To obtain the second line, we used the relation $\beta \sqrt{\gamma'} = \beta' \sqrt{\gamma}$ and the fact that the parameters are independent of the imaginary time. 

Next, we consider the variation of $\psi$ with respect to  $\bar{t}$, which changes the hypersurface and $n_\mu$. 
We obtain
\begin{equation}
\begin{split}
d\psi =d(\beta'p)  = p_\mu d\beta^\mu + n' d\nu -\beta^\mu pdn_\mu
\,. \end{split}
\end{equation}
By using this relation, the time derivative of $\Psi^{(0)}[\pt,\lambda]$ reads
\begin{equation}
\begin{split}
\partial_\pt\Psi^{(0)}[\pt,\lambda]
 & = \int d\Sigma_{\pt}N\covDer_\mu  (\beta^\mu p)\\
 & = \int d\Sigma_{\pt}N\Bigl[
 (\covDer_\mu \beta^\mu)p
 +  
 \beta^\mu(\covDer_\mu
  \frac{1}{\beta'})\beta'p
 +   \frac{\beta^\mu}{\beta'} \covDer_\mu(\beta'p)
 \Bigr]
 \\
 & = \int d\Sigma_{\pt}N\Bigl[
 (\covDer_\mu \beta^\mu)p
 +  \frac{\beta^\nu}{\beta'}({n_\mu \covDer_\nu\beta^\mu +\beta^\mu \covDer_\nu n_\mu})p\\
&\qquad +   \frac{\beta^\nu}{\beta'} (p_\mu \covDer_\nu \beta^\mu +n' \covDer_\nu \nu -p\beta^\mu (\covDer_\nu n_\mu))
 \Bigr]
 \\
 & = \int d\Sigma_{\pt}N\Bigl[
 (\covDer_\mu \beta^\nu)\Bigl( \delta^\mu_\nu p
  +    \frac{\beta^\mu}{\beta'} (p_\nu+n_\nu p) \Bigr)
 +\frac{\beta^\mu}{\beta'} n' \covDer_\mu\nu
 \Bigr].
\label{eq:derPhi2}
\end{split}
\end{equation}
Comparing Eq.~\eqref{eq:delPsi} with Eq.~\eqref{eq:derPhi2}, 
we obtain the expectation values of the energy-momentum tensor and the particle current by the local Gibbs distribution as
\begin{align}
\averageLG{ {\hT^{\mu}_{~\nu}}(x)}_\pt&=\delta^\mu_\nu p
  +    \frac{\beta^\mu}{\beta'} (p_\nu+n_\nu p)
\notag \\
&=  (e+p)u^\mu u_\nu + \delta^\mu_\nu p
, \label{eq:perfect1}\\
\averageLG{\hat{J}^{\nu}(x)}_\pt &=  n'\frac{\beta^\mu}{\beta'} =  n u^\mu, \label{eq:perfect2}
\end{align}
where
$e\equiv \averageLG{ {\hT^{\mu}_{~\nu}}(x)}_\pt u_\mu u^\nu =  p_\nu u^\nu \beta^\mu u_\mu/\beta' $,
and $n=-n' \beta^\mu u_\mu/\beta'$. 
Here we used that the energy-momentum tensor is symmetric under the change of the indices to derive the 
second line in Eq.~\eqref{eq:perfect1}.
Equations~\eqref{eq:perfect1} and~\eqref{eq:perfect2}  are nothing but the constitutive relations of the energy-momentum tensor and the particle current in a perfect fluid.

\subsubsection{First order: Navier-Stokes equations} \label{sec:FirstOrder}
Let us consider the next leading order in the derivative expansion.
We need not consider the derivative corrections coming from $\averageLG{\hcurrent_a^\mu(x)}_\pt$, since $\Psi^{(1)}$ vanishes
for the parity-symmetric system. 
The first-order correction to the dissipative part comes from
\begin{equation}
\begin{split}
\int_0^{1}d\tau \averageLG{T_\tau \hSigma_\tau(\pt,\pt_0) \delQ\hcurrent_a^\mu (x)}_{\pt} 
=\bm{(}\delQ\hcurrent_a^\mu (x), \hSigma(\pt,\pt_0)\bm{)}_\pt .
\end{split}
\end{equation}
We used the inner product Eq.~\eqref{eq:innerProd} and the Hermite symmetry of the inner product.
Then, the first-order corrections read
\begin{align}
\average{\delQ\hT^{\mu\nu}(x)}_{(0,1)}
&\simeq h^{\mu\nu}\bm{(} \delQ\hp(x), \hSigma(\pt,\pt_0)\bm{)}_\pt +\bm{(} \delQ\hpi^{\mu\nu}(x), \hSigma(\pt,\pt_0)\bm{)}_\pt, \label{eq:delT01}\\
\average{\delQ\hJ^{\mu}(x)}_{(0,1)}
&\simeq \bm{(} \delQ\hJ^{\mu}(x), \hSigma(\pt,\pt_0)\bm{)}_\pt \label{eq:delJ01}.
\end{align}
where $\simeq$ denotes an equality at the first order in derivatives.
The right-hand side of Eqs.~\eqref{eq:delT01} and \eqref{eq:delJ01} also contain the higher-order contributions.
In the first order in the derivative expansion, we can neglect $\delta\hlambda^a\covDer_\mu\average{\delta\hcurrent_a^\mu}$ in 
$\hSigma(\pt,\pt_0)$ because $\average{\delQ\hcurrent_a^\mu}= O(\covDer )$ and thus $\covDer_\mu\average{\delQ\hcurrent_a^\mu}=O(\covDer^2)$.
We can replace $\hK$ in these inner products with $\hat{P}_\mu \beta^\mu(x)$. 
We remark here that the dissipative corrections are orthogonal to $n_\mu$ by construction, 
and thus we do not need to employ $n^\mu$ or $v^\nu$ for the tensor decomposition.
Therefore, we may decompose these inner products in Eqs.~\eqref{eq:delT01} and \eqref{eq:delJ01} by only using $h^{\mu\nu}$.
Two-point correlation functions with odd numbers of indices, such as 
$\bm{(} \delQ\hat{p}(x), \delQ\hJ^\mu(x')\bm{)}_\pt$, vanish. 
Furthermore, correlation functions with a single $\delQ\hpi^{\mu\nu}(x)$ also vanish since $\delQ\hpi^{\mu\nu}(x)$ is traceless.
In consequence, we have
\begin{align}
\bm{(} \delQ\hat{p}(x), \hSigma(\pt,\pt_0)\bm{)}_\pt
&=
- {\int_{\pt_0}^\pt d\ptc{t}' \int d\Sigma_{\ptc{t}'} N'} \bm{(} \delQ\hat{p}(x),  \delQ \hat{p}(x')\bm{)}_\pt h^{\mu\nu}(x')\covDer_{\mu}\beta_{\nu}(x') 
\notag \\
&\simeq
-\frac{\zeta}{\beta(x)} h^{\mu\nu}(x)\covDer_{\mu}\beta_{\nu}(x),\\
\bm{(} \delQ\hat{\pi}^{\mu\nu}(x), \hSigma(\pt,\pt_0)\bm{)}_\pt&=
-{\int_{\pt_0}^\pt d\ptc{t}' \int d\Sigma_{\ptc{t}'} N'}\bm{(} \delQ\hat{\pi}^{\mu\nu}(x),  \delQ \hat{\pi}^{\rho\sigma}(x')\bm{)}_\pt \covDer_{\langle\rho}\beta_{\sigma\rangle}(x')
\notag \\
&\simeq -\frac{2\eta }{\beta(x)}h^{\mu\rho}(x)h^{\nu\sigma}(x) \covDer_{\langle\rho}\beta_{\sigma\rangle}(x),\\
\bm{(} \delQ \hJ^{\mu}(x), \hSigma(\pt,\pt_0)\bm{)}_\pt&=
-{\int_{\pt_0}^\pt d\ptc{t}' \int d\Sigma_{\ptc{t}'} N'}\bm{(} \delQ\hJ^{\mu}(x),
\delQ \hJ^\nu(x')\bm{)}_\pt \covDer_{\perp\nu}\nu(x')
\notag \\
& \simeq-\frac{\kappa}{\beta(x)} \covDer_{\perp}^\mu\nu(x) ,
\end{align}
where we used $\partial_\mu\lambda^a(x')\simeq \partial_\mu\lambda^a(x)$.
Here the transport coefficients, $\zeta$, $\eta$, and $\kappa$, are the bulk viscosity, the shear viscosity, and the diffusion constant, respectively. 
They are given by the Kubo formulas:
\begin{align}
\zeta &=\beta(x){\int_{-\infty}^\pt d\ptc{t}' \int d\Sigma_{\ptc{t}'} N'}\bm{(}\delQ\hp(x'),\delQ \hp(x) \bm{)}_\pt, 
\label{eq:bulk}\\
\eta &=\frac{\beta(x)}{(d+1)(d-2)}{\int_{-\infty}^\pt d\ptc{t}' \int d\Sigma_{\ptc{t}'} N'}\bm{(}\delQ\hat{\pi}^{\mu\nu}(x'), \delQ\hat{\pi}^{\rho\sigma}(x)\bm{)}_\pt h_{\mu\rho}(x)h_{\nu\sigma}(x) ,
\label{eq:shear}\\
\kappa &=\frac{\beta(x)}{d-1}{\int_{-\infty}^\pt d\ptc{t}' \int d\Sigma_{\ptc{t}'} N'}\bm{(}\delQ\hat{J}^\mu(x'), \delQ\hat{J}^\nu(x)\bm{)}_\pt h_{\mu\nu}(x),
\label{eq:diffusion}
\end{align}
where we replaced $\pt_0$ by $-\infty$, which can be justified in the first order in the derivative expansion.
We can now construct the constitutive relations up to the first order,
which are given as Eqs. \eqref{eq:delTmunu}, and \eqref{eq:delJmu}.
Once we calculate the transport coefficients, $\zeta$, $\eta$, $\kappa$, and the pressure $p(\beta,\nu)$ from the microscopic theory, 
we have closed equations composed of the continuity equations.
These are nothing but relativistic versions of the Navier-Stokes equations. 
We emphasize here that we derive them without choosing a frame such as the Landau-Lifshitz or Eckart frame. 

\subsection{Choice of frame} \label{sec:FrameChoise}
In relativistic hydrodynamics, we face the frame ambiguity,
which stems from a way to define the fluid four-velocity.
One useful frame is the Landau-Lifshitz frame, 
in which the energy flux of a fluid element vanishes at the rest frame of the fluid. 
Another is the Eckart frame, in which the particle flux is absent.
In our approach, the choice of $v^\mu$ and $n_\mu$ corresponds to the choice of frames. 
In this subsection, we show that by explicitly choosing $v^\mu$ and $n^{\mu}$, 
our constitutive relations reproduce
those in the Landau-Lifshitz and Eckart frames within the derivative expansion.

\subsubsection{Landau-Lifshitz frame}

The fluid four-velocity in the Landau-Lifshitz frame is defined by the condition 
that in the local rest frame, the energy flux of a fluid element vanishes. 
Then, the energy and charge densities coincide with the local thermodynamic values.
In other words, the Landau-Lifshitz frame is defined by~\cite{Landau:Fluid} 
\begin{equation}
 \average{\delta\hT ^{\mu\nu}(x)}u_{L\nu}(x) = 0, \quad \average{\delta\hJ^\mu (x)}u_{L\mu}(x) = 0, \label{eq:LandauCondition}
\end{equation}
where the subscript $L$ denotes the Landau-Lifshitz frame. 
We can easily see that Eq.~\eqref{eq:LandauCondition} is satisfied if we choose $u_L^\mu\equiv v^\mu =n^\mu=u^\mu $.
In this case, we have a familiar projection
$h^{\mu\nu} = g^{\mu\nu}+u_L^\mu u_L^\nu $. 
The constitutive relations up to first order in the derivative expansion read
\begin{align}
  \average{\hT^{\mu\nu}(x)}&= (e+p) u_L^\mu u_L^\nu + p g^{\mu\nu}-2\eta \sigma^{\mu\nu} - \zeta \theta h^{\mu\nu},
   \label{eq:Landau1} \\
  \average{\hJ^{\mu}(x)}&= nu_L^\mu -{\frac{\kappa}{\beta}} \covDer_\perp^\mu \nu ,
 \label{eq:Landau2}
\end{align}
where 
\begin{equation}
\begin{split}
\sigma^{\mu\nu} \equiv 
 \frac{1}{2}h^{\mu\alpha}h^{\nu\beta}(\covDer_{\alpha}u_{L\beta}+\covDer_{\beta}u_{L\alpha})-\frac{1}{d-1}h^{\mu\nu}h^{\alpha\beta}\covDer_{\alpha}u_{L\beta},
  \quad\theta \equiv \covDer_\mu u_L^\mu. \label{eq:sigma}
\end{split}
\end{equation}

In this frame, we can explicitly write down the projected operators in Eqs.~(\ref{eq:bulk})-(\ref{eq:diffusion}) as
\begin{align}
\delQ \hp&= \delta\hp-\Bigl(\frac{\partial p}{\partial n}\Bigr)_e \delta\hn - \Bigl(\frac{\partial p}{\partial e}\Bigr)_n\delta\he,\\
\delQ \hpi^{\mu\nu}&= \delta\hpi^{\mu\nu},\\
\delQ \hat{J}^\mu&= \delta\hJ^\mu-  \frac{n}{e+p}h^{\mu\nu}\delta\hp_{\nu}.
\end{align}
To derive these equations, we used 
\begin{equation}
\begin{split}
\hProj\delta\hp  = \int d\Sigma'_\pt \delta\hc_a(x')\frac{\delta }{\delta c_a(x')}\averageLG{\hp(x)}_\pt = \Bigl(\frac{\partial p}{\partial n}\Bigr)_e \delta\hn + \Bigl(\frac{\partial p}{\partial e}\Bigr)_n\delta\he + O(\covDer_\perp),
\end{split}
\end{equation}
\begin{equation}
\begin{split}
\hProj \delta\hJ^\mu =  \int d\Sigma_\pt  \int d\Sigma'_\pt \delta\hp_\rho(x) \bm{(}\delta\hp_\rho(x),\delta\hp_\nu(x') \bm{)}_\pt^{-1} \bm{(}\delta\hp_\nu(x'), \delta\hJ^\mu \bm{)}_\pt
=h^{\mu\nu}\delta\hp_{\nu} \frac{n}{e+p}+ O(\covDer_\perp),
\end{split}
\end{equation}
where $\delta\he\equiv -u_L^\mu \delta \hp_\mu$, 
and we used the following relations~\cite{Minami:2012hs}:
\begin{align}
\int d\Sigma_\pt \bm{(}\delta\hp_\rho(x),\delta\hp_\nu(x') \bm{)}_\pt &= \frac{1}{\beta}h_{\rho\nu}(e+p)+ O(\covDer_\perp),\\
\int d\Sigma_\pt \bm{(}\delta\hp_\nu(x),\delta\hJ^\mu(x') \bm{)}_\pt &= \frac{n}{\beta}P_\nu^\mu+ O(\covDer_\perp).
\end{align}

\subsubsection{Eckart frame}
Next, we consider the Eckart frame. 
The fluid four-velocity for the Eckart frame is defined by the condition 
that it is proportional to the particle current, i.e., $u^\mu_E(x) \equiv J^\mu(x)/\sqrt{-J^\mu(x) J_\mu(x)}$, 
where the subscript $E$ denotes the Eckart frame, and $J^\mu(x) = \average{\hJ^\mu(x)}$~\cite{Eckart:1940te}.
It is also required that the energy density is expressed as $ e=u^E_\mu \average{\hT ^{\mu\nu}(x)} u^E_\nu(x)$.
In the first order in the derivative expansion, we may choose $v^\mu$ and $n^\mu$ as 
\begin{equation}
\begin{split}
v^\mu=n^\mu =u_E^\mu= \frac{1}{\sqrt{-\left(u^\mu -\frac{\kappa}{\beta n}\covDer_{\perp}^\mu\nu \right)^2}} \left(u^\mu -\frac{\kappa}{\beta n}\covDer_{\perp}^\mu\nu \right)  = u^\mu -\frac{\kappa}{{\beta} n}\covDer_{\perp}^\mu\nu +O(\covDer_\perp^2) .
\label{eq:Eckart}
\end{split}
\end{equation}
Using $u^\mu=u_E^\mu+ (\kappa/({\beta} n))\partial_\perp^\mu \nu+O(\covDer^2)$, we obtain
\begin{align}
  \average{\hT^{\mu\nu}(x)}&= (e+p) u_E^\mu u_E^\nu + p g^{\mu\nu} 
  + q^\mu u_E^\nu + u_E^\mu q^\nu  -2\eta \sigma^{\mu\nu} - \zeta \theta h^{\mu\nu},\\
  \average{\hJ^{\mu}(x)}&= nu_E^\mu,
\end{align}
where we dropped the terms of order $\covDer^2_\perp$. $\sigma^{\mu\nu} $ and $\theta$ are obtained by replacing $u^\mu_L$ in Eq.~\eqref{eq:sigma} with $u^\mu_E$.
The thermal conductivity $q^\mu$, which is absent in the Landau-Lifshitz frame reads
\begin{equation}
 \begin{split}
  q^\mu = \frac{e+p}{n{\beta}}\kappa \covDer_\perp^\mu \nu.
  \label{eq:Eckart-like}
 \end{split}
\end{equation}
We note that the shear and bulk viscous terms are 
the same as those of the Landau-Lifshitz frame. 

Although we do not have the charge diffusion in this frame, 
the expression of heat current is slightly different from the original Eckart one $q_E^\mu$, which is given by \cite{Eckart:1940te}
\begin{equation}
 \begin{split}
  q_{E}^\mu = -\lambda ( \covDer_\perp^\mu T + T \delt u_E ^\mu ) ,
 \end{split}
\end{equation}
where $\lambda$ denotes the thermal conductivity of the fluid.
The apparent difference is coming from whether we use the time derivative of the fluid four-velocity in order to construct the constitutive relations.
Although we utilize the Mori projection operator to eliminate the time derivative
of the parameters from the entropy production, we can reconstruct the constitutive relations by using the time derivative terms with the help of the equation of motion. 
In the first order, 
we can use the equation of motion for the perfect fluid,
\begin{equation}
 \begin{split}
  \delt u^{\mu} = -\frac{1}{T}\covDer_{\perp}^{\mu} T - \frac{nT}{e+p} \covDer_{\perp}^{\mu} \nu ,
  \end{split}
\end{equation}
in order to eliminate $\covDer_{\perp}^{\mu} \nu$ from Eq.~\eqref{eq:Eckart-like}.
Then, we derive the constitutive relations in the original Eckart frame 
with $\lambda = ((e+p)^2{\beta}/n^2) \kappa$.

Obviously, in our formalism, the constitutive relations in the Landau-Lifshitz and Eckart frames are equivalent within the first order in the derivative expansion.
These are related to each other by the redefinition of the fluid four-velocity, 
$u_L^\mu \leftrightarrow u_E^\mu + (\kappa/(\beta n))\covDer_\perp^\mu \nu$ in Eqs.~\eqref{eq:Landau1} and \eqref{eq:Landau2}.
More generally, if we choose a frame such that $v^\mu=u^\mu+O(\nabla)$ and $n^\mu=u^\mu+O(\nabla)$, 
the constitutive relations in this frame are equivalent to those in the Landau frame
within the first order in the derivative expansion. Namely, if $n^\mu$ is a functional of $\lambda^a$, 
the constitutive relations are unique and become those in the Landau-Lifshitz frame. 
We note that such a uniqueness was also discussed in Ref.~\cite{PhysRevE.87.053008} based on the Boltzmann equation.

\section{Discussion}\label{sec:Discussion}
In this paper, we have derived hydrodynamic equations from quantum field
theory by assuming that the density operator has the form of the local
Gibbs distribution at initial time.
In particular, we have derived the first-order equations, that is, the relativistic version of the Navier-Stokes equation 
without a choice of frames such as the Landau-Lifshitz or Eckart frame.
Our frame-independent analysis becomes important if the vorticity is nonzero; in this case, we cannot choose $n_\mu=u_\mu$ because the 
vorticity, constructed from $n_\mu$ and $ n_\nu\epsilon^{\mu\nu\rho\sigma}\covDer_\rho n_\sigma$, vanishes by the Frobenius theorem~\cite{Becattini:2014yxa}. 

\begin{figure}[h]
\includegraphics[width=0.75\linewidth]{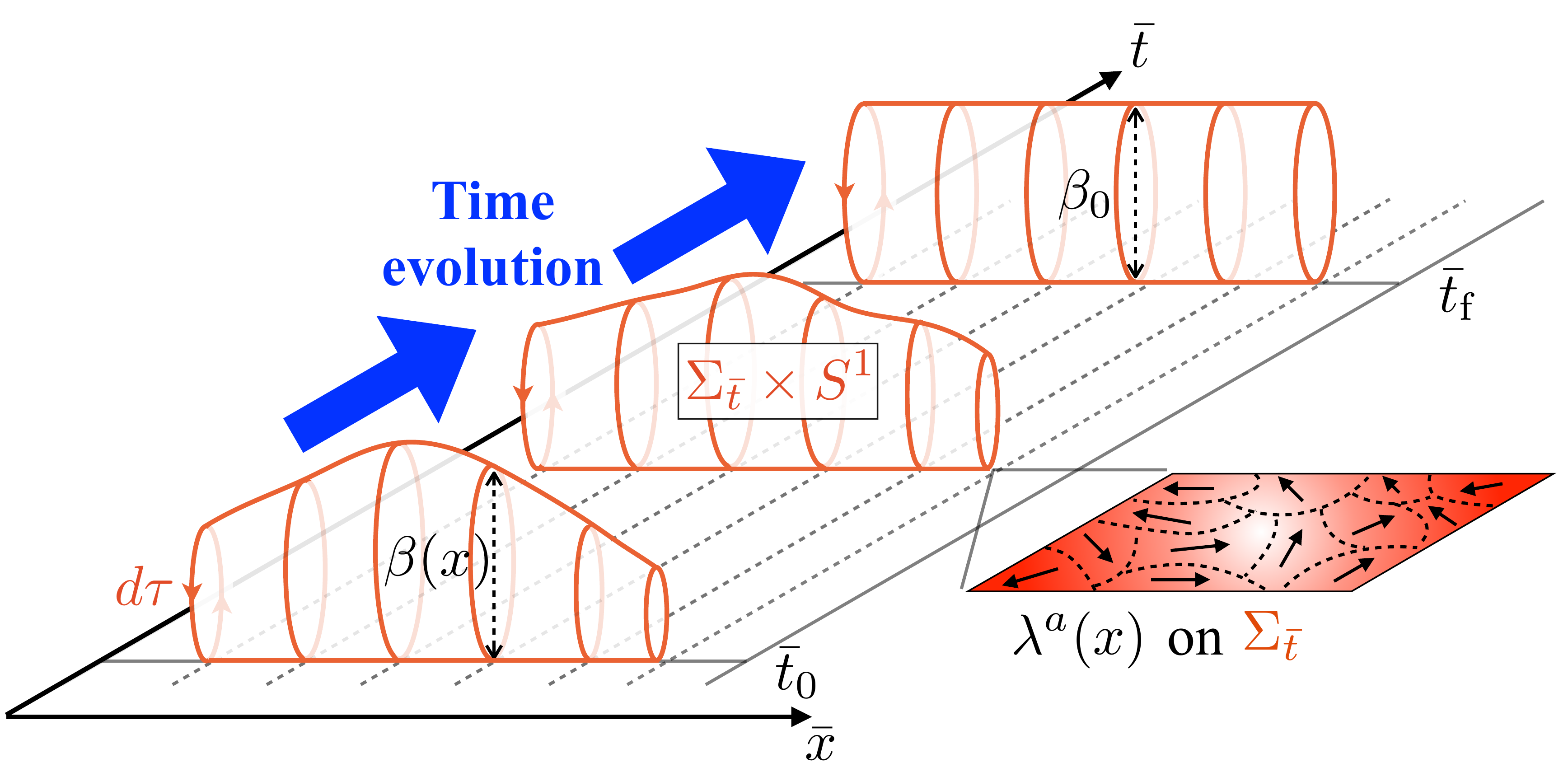}
\caption{
Schematic figure of the real-time evolution in our formulation toward the global thermal equilibrium. 
} 
\label{Fig:time-evolution}
\end{figure}

The real-time evolution in our formulation is schematically shown in
Fig.~\ref{Fig:time-evolution}. 
The density operator of the system at initial time $\pt_0$ is assumed to have the form of the local Gibbs distribution. 
Then we expand the density operator at a later time $\bar{t}$ around the new local Gibbs distribution with the thermodynamic parameters $\lambda^a(x)$ at that time. 
In each time, the local Gibbs distribution (the Massieu-Planck
functional) can be expressed 
by using the imaginary-time path integral under the curved spacetime background $\Sigma_\pt \times S^1$, whose metric is given in Eq.~\eqref{eq:thermalMetric}. 
After a sufficiently long time, the system reaches the global thermal equilibrium with the uniform imaginary-time radius $\beta_0$.
The local Gibbs distribution enables us to treat a nonequilibrium state beyond the real-time formalism~\cite{le2000thermal}, 
in which the distribution is necessarily in the global equilibrium.
However, in an early stage far from equilibrium, the density operator cannot be approximated by the local Gibbs distribution, 
and thus our formulation is no longer applicable.

As mentioned in the Introduction, our method is closely related to that presented by Sasa~\cite{sasa:2013}.
In fact, if we take $n_\mu=(-1,\bm{0})$ in the flat spacetime, they are equivalent.
The difference is that our formalism is based on the Heisenberg picture, 
while that in Ref.~\cite{sasa:2013} is based on the Schr\"odinger one; 
these are related to each other by the unitary transformation, $\hrho^{\text{LG,Sasa}}_t  = e^{i\hH t} \hrho^{\text{LG}}_t  e^{-i\hH t}$.

There are several directions on future research based on this method:
One is the generalization to a system with a quantum anomaly such as chiral fermions in which the matter couples to external gauge fields.
This generalization is straightforward: We may replace the energy-momentum tensors and the particle current to those in background gauge fields.
In this case, $\hK[\pt;\lambda]$ formally has the same form as before.
The difference is that  the currents are no longer conserved,
\begin{align}
\covDer_\mu \hT^{\mu\nu} &= F_{\mu\nu}\hJ^{\mu} , \label{eq:ExternalfieldConservationLaw1}\\
\covDer_\mu \hJ^{\mu} &= C_\text{ano}\epsilon^{\mu\nu\rho\sigma}F_{\mu\nu}{F}_{\rho\sigma}, \label{eq:ExternalfieldConservationLaw2}
\end{align}
where $F_{\mu\nu}$ is the field strength of the external gauge field,  $\epsilon^{\mu\nu\rho\sigma}$ the antisymmetric tensor, and
$C_\text{ano}$ the anomaly coefficient.
Using Eqs.~(\ref{eq:ExternalfieldConservationLaw1}) and (\ref{eq:ExternalfieldConservationLaw2}) instead of Eqs.~(\ref{eq:conservation2}) and (\ref{eq:conservation1}), the divergence of the entropy operator reads as
$\covDer_\mu \hs^\mu =
 - (\covDer_\nu\beta^\mu) \delta {\hT^{\mu}}_{~{\nu}}  -\delta \hJ^{\mu}(x)f_\mu $,
where $f_\mu \equiv \covDer_\mu \nu
+\beta^\nu  F_{\mu\nu}$. 
The term $ \nu C_\text{ano}\epsilon^{\mu\nu\rho\sigma}F_{\mu\nu}{F}_{\rho\sigma}$ coming from the anomaly cancels out in the divergence of the entropy current operator.
Therefore, the anomaly does not directly contribute to the dissipative part of the currents,
which is consistent with the observation in the entropy-production method~\cite{Son:2009tf} and the generating-functional method~\cite{Banerjee:2012iz,Jensen:2012jh}.

Another direction is  an application to second-order hydrodynamic equations.
There are several works derivingthese equations from microscopic theories~\cite{Muronga:2006zx,Tsumura:2007ji,*Tsumura:2011cj,York:2008rr,Betz:2008me,Monnai:2009ad,Van:2011yn,Denicol:2012cn,Jaiswal:2013npa}, which are based on the Boltzmann equation.
 In contrast to them, our method is applicable to strongly coupled systems.
In general, all possible terms respecting symmetries appear in the derivative expansion, whose coefficients depend on details of the system.
Our method gives the Kubo formulas for these coefficients.
We may obtain Kubo formulas different from those in the analyses based on the Boltzmann equation.
We leave these interesting applications for future work.

\acknowledgements
We thank Yuki Minami for collaboration in the early stage of this work.
We also thank  M.~Fukuma, T.~Kunihiro, S.~Sasa, and Y.~Tanizaki for useful discussions.  
T.~H. was supported by a JSPS Research Fellowships for Young Scientists (Grant No. 24008301).
Y.~H. was supported by JSPS KAKENHI (Grants No. 24740184).
M.~H. was supported by the RIKEN Junior Research Associate Program.
T.~N. was supported by the Special Postdoctoral Researchers Program at RIKEN.
This work was partially supported  by the RIKEN iTHES Project.

\appendix
\section{Derivation of Eq.~\eqref{eq:StokesTheorem}}\label{sec:StokesTheorem}
Let us, here derive  Eq.~\eqref{eq:StokesTheorem}.
Noting that the volume element can be written as $d\Sigma_{\pt\mu}=d^dx\sqrt{-g}\delta(\pt-\timefunc(x)) \partial_\mu  \timefunc(x)= -d^dx\sqrt{-g} \partial_\mu\theta(\pt-\timefunc(x)) $,
we write
\begin{equation}
\begin{split}
\int d\Sigma_{\pt\mu}  f^{\mu}(x) &= 
-\int d^dx \sqrt{-g} \partial_\mu\theta(\pt-\timefunc(x))   f^{\mu}(x) \\
& = 
\int d^dx \sqrt{-g}\theta(\pt-\timefunc(x))   \covDer_\mu f^{\mu}(x), \label{eq:Stokes1}
\end{split}
\end{equation}
where we used the integral by part, and  assumed that $f^\mu(x)$ vanishes at the boundary.
The derivative of Eq.~\eqref{eq:Stokes1} with respect to $\pt$ leads to Eq.~\eqref{eq:StokesTheorem},
\begin{equation}
\begin{split}
\partial_{\pt} \int d\Sigma_{\pt\mu}  f^{\mu}(x) &= 
\partial_{\pt} \int d^dx\sqrt{-g} \theta(\pt-\timefunc(x))   \covDer_\mu f^{\mu}(x) \\
& = 
 \int d\Sigma_\pt N(x)   \covDer_\mu f^{\mu}(x),
 \label{eq:Stokes}
\end{split}
\end{equation}
where we used $d\Sigma_\pt=d^dx\sqrt{-g}\delta(\pt-\timefunc(x))N^{-1}$.

\bibliography{hydro}
\end{document}